\documentclass[namedreferences,hyperref,optionalrh,solaromanenum]{spr-sola}

\usepackage{graphicx}                    
\usepackage{color}                       
\usepackage{amsmath}
\usepackage{graphicx}
\usepackage{txfonts}
\usepackage{hyperref}
\usepackage{textcomp}
\usepackage{gensymb}
\DeclareMathOperator{\diverg}{div}

\newcommand{\fdg}{.\!\!\degree}

\chardef\us=`\_

\begin{document}

\begin{frontmatter}

\title{Activity-Cycle Variations of Convection Scales in Subsurface Layers of the Sun}

%
\author[addressref={aff1},corref,email={A.Getling@mail.ru}]{\inits{A.V.}\fnm{Alexander V. }\snm{Getling}\orcid{0000-0002-1766-2620}}
\author[addressref={aff2},email={alexander.g.kosovichev@njit.edu}]{\inits{A.G.}\fnm{Alexander G. }\snm{Kosovichev}\orcid{0000-0003-0364-4883}}

\address[id=aff1]{Skobeltsyn Institute of Nuclear Physics, Lomonosov Moscow State
		University, Moscow, 119991 Russia}
\address[id=aff2]{New Jersey Institute of Technology, NJ 07102, USA}

%
\runningauthor{Getling and Kosovichev}
\runningtitle{\textit{Solar Physics} Variation of the solar-convection scales}


\begin{abstract}
We use subsurface-flow velocity maps inferred by time--distance helioseismology from Doppler measurements   
with the Helioseismic and Magnetic Imager (HMI) of the Solar Dynamics Observatory (SDO) to investigate variations of large-scale convection during Solar Cycles 24 and 25 in the 19-Mm-deep layer. The spatial power spectra of the horizontal-flow divergence reveal well-defined characteristic scales of solar supergranulation in the upper 4~Mm layer, while the giant-cell scale is prominent below levels of $d \sim 8$~Mm. We find that the characteristic scales of supergranulation remain stable while the giant scales increase during the periods of the 11-year activity cycle maxima. The power of the giant-cell scales increases with the enhancement of solar activity. This may be due to large-scale flows around active regions and, presumably, solar-cycle variations of the convection-zone stratification.
\end{abstract}

%

\end{frontmatter}

%
	\section{Introduction} \label{sec:intro}
	
	Solar convection is a crucial component of the dynamo processes that generate magnetic fields, which govern the active phenomena on the Sun. That is why investigations of the structure and dynamics of solar convective flows are of primary importance for understanding solar activity mechanisms. In this respect, the spatial and temporal variations of the characteristic scales of large-scale convection are of much interest. Particular attention should be given to their variations during the 11-year solar cycle, as they can be related to the factors that determine the cyclic behavior of solar activity. Observations of the solar surface reveal two primary large convective scales---supergranulation and giant cells \citep{Rincon_Rieutord_2018,Hathaway_etal_2013}.
    
    Methods of local helioseismology---ring-diagram analysis and time-distance helio\-seismology---provide information about the large-scale convection in subsurface layers, covering the top 20~Mm of the convection zone. This layer is characterized by a strong radial gradient in the solar rotation rate and is known as the Near-Surface Shear Layer (NSSL). Our previous investigation of the convection spectrum in this layer using the flow velocity maps obtained by the time--distance helioseismology technique and available from the SDO Joint Science Operations Center \citep[JSOC; see][]{Zhao_2012} showed the convective motions on the supergranulation scale ($\sim 20$--30~Mm) to be dominant in the upper 4-Mm layer and the giant-cell scales ($\sim 100$--200~Mm) to become prominent below 8~Mm \citep{Getling_Kosovichev_2022}. 

        For a comprehensive review of supergranulation properties, we refer the reader to an article by \cite{Rincon_Rieutord_2018}. Short-term variations in supergranule sizes and velocities were analyzed by \citet{Williams_etal_2014}. In particular, \cite{Williams_Pesnell_2014} detected fluctuations with periods of three to five days. In the context of our study, supergranular-scale variations in the solar activity cycle are particularly interesting. Thus, it is worth mentioning a study by \cite{Raju_Singh_2002}, who found that enhanced network magnetic fields reduce the supergranular cell size. \citet{McIntosh_etal_2011} have demonstrated that the variation of the supergranular length scale during periods of low solar activity affects the spectral irradiance of the Sun. According to \cite{Tlatov_2012}, the supergranule size exhibits a positive correlation with solar activity, reaching a maximum in about 1.5 years after the activity maximum. \cite{Mandal_etal_2017} found that the mean supergranular scale is correlated with the amplitude of the sunspot-number cycle and with the total solar irradiance. \cite{Chatterjee_etal_2017} studied solar-cycle variations of a number of supergranule parameters using a one-century-long dataset. In particular, they found that the supergranular scale in active regions varies coherently with the sunspot number (a correlation coefficient of 0.90) and is anticorrelated with it in quiet regions (a correlation coefficient of $-0.86$). 
    
It should be noted that large-scale structures in the subsurface layers of the Sun are not exhausted by convection cells. Numerous toroidal (involving only horizontal motions) oscillation modes with periods of order the solar rotational period and spatial scales of order the solar radius have been detected \citep[see, e.g.,][]{Gizon_etal_2021,Mandal2024}. Such spatial scales suggest that the Coriolis force plays a fundamental role in their dynamics, serving as their restoring force. Such modes came to be known as inertial modes \citep[among them, the Rossby mode well known in geophysical fluid dynamics was predicted for slowly rotating white dwarfs by][]{Papaloizou_Pringle_1978}. In particular, \cite{Gizon_etal_2021} suggested that large-scale features of the velocity field revealed at high latitudes \citep[][and references therein]{Hathaway_Upton_2021} are manifestations of inertial modes rather than giant convection cells. There are also other giant-scale modes, detectable at lower latitudes \citep{Hanson_Hanasoge_2024}. More research is needed to make a definite choice between inertial modes and giant convection cells. Such an investigation falls outside the scope of this study, and we will refer to the giant structures under consideration as convection cells without further stipulation, leaving the solution of the aforementioned dilemma for future studies.

We study here the solar-cycle variation of the characteristic scales of large-scale convection, i.e., supergranulation and giant cells, using time-distance helioseismology data from the SDO/HMI instrument and calculating the spatial power spectrum of solar convection at various depths. Specifically, we define the value of the spherical-harmonic degree corresponding to the characteristic convection wavelength as the degree value corresponding to the center of mass of the power spectrum calculated by spherical-harmonic decomposition of the horizontal-flow divergence and investigate how this characteristic scale varies with depth during the period of the SDO observations covering Solar Cycle 24 and the first half of Cycle 25.

    \section{The Data and the Processing Techniques Used}\label{obs}
	
	The data of helioseismological determinations of the azimuthal and meridional velocity components, $v_x$ and $v_y$, used in our analysis are available from the SDO JSOC data analysis pipeline \citep{Zhao_2012}.  The subsurface flow maps for the whole visible hemisphere of the Sun are routinely produced every 8 hours for a grid of $1026\times 1026$ points spanning over 123\degree\ of heliographic colatitude, $\theta$, and Stonyhurst longitude, $\varphi$, with a spatial resolution of 0\fdg 12. The flow maps are produced for the following eight characteristic depths below the photosphere, $d$ (the corresponding depth ranges for which the inversions were done are parenthesized): 0) $d=0.50$ (0--1)~Mm, 1) $d=2.00$ (1--3)~Mm, 2) $d=4.00$ (3--5)~Mm, 3) $d=6.00$ (5--7)~Mm, 4) $d=8.50$ (7--10)~Mm, 5) $d=11.50$ (10--13)~Mm, 6) $d=15.0$ (13--17)~Mm, 7) $d=19.0$ (17--21)~Mm.
	
	The period from May 2010 to August 2024 will be considered. We will need the spatial spectra of the convective horizontal-velocity field at different depths. For our spectral analysis, a scalar function, viz., the divergence of the horizontal velocity, 
	\begin{equation}
		f(\theta,\varphi)=\diverg \mathbf V(\theta,\varphi),
	\end{equation} is more convenient than the velocity vector $\mathbf V(\theta,\varphi)$ itself. We represent $f(\theta,\varphi)$ as a spectral decomposition
	\begin{equation}
		f(\theta,\varphi) = \sum_{\ell=0}^{\ell_{\max}}\sum_{m=-\ell}^{\ell}A_{\ell m}Y_{\ell}^{m}(\theta,\varphi)
		\label{series}
	\end{equation}
	in spherical harmonics of degree $\ell$ and azimuthal number $m$
	(here, $\ell_{\max}$ is a suitably chosen upper spectral boundary). We are interested in the power spectrum of the flow,
	\begin{equation}\label{power}
		p_{\ell m}=|A_{\ell m}|^2.
	\end{equation}
   According to Parseval's theorem, the integrated power of the flow described by the function $f(\theta,\varphi)$ can be written as
	\begin{equation}
		p_\mathrm{tot} \equiv \int\limits_\Omega f^2 \mathrm d\Omega =\sum_{\ell=0}^\infty \sum_{m=-\ell}^\ell |A_{\ell m}|^2
	\end{equation}
	(here, $\mathrm d\Omega$ is an elementary and $\Omega$ is the full solid angle).
 
    The wavelength of the harmonic  $Y_{\ell}^{m}(\theta,\varphi)$ on a sphere of radius $r$ is determined by the \cite{Jeans_1923} formula
	\begin{equation}
		\lambda=\frac{2\pi r}{\sqrt{\ell(\ell+1)}}.
		\label{Jeanseq}
	\end{equation}
We use it to estimate the characteristic scales of convection structures, assuming that $r=R_\odot-d$,  where $R_\odot$ is the solar radius (however, for our order-of-magnitude estimates, $r$ can be set equal to $R_\odot$). The wavelength $\lambda$ approximately determines the characteristic convection scale, which depends on the configuration of the convection structures.

    \begin{figure*}
	\centering{	\includegraphics[width=0.45\textwidth,bb=20 0 780 650, clip]{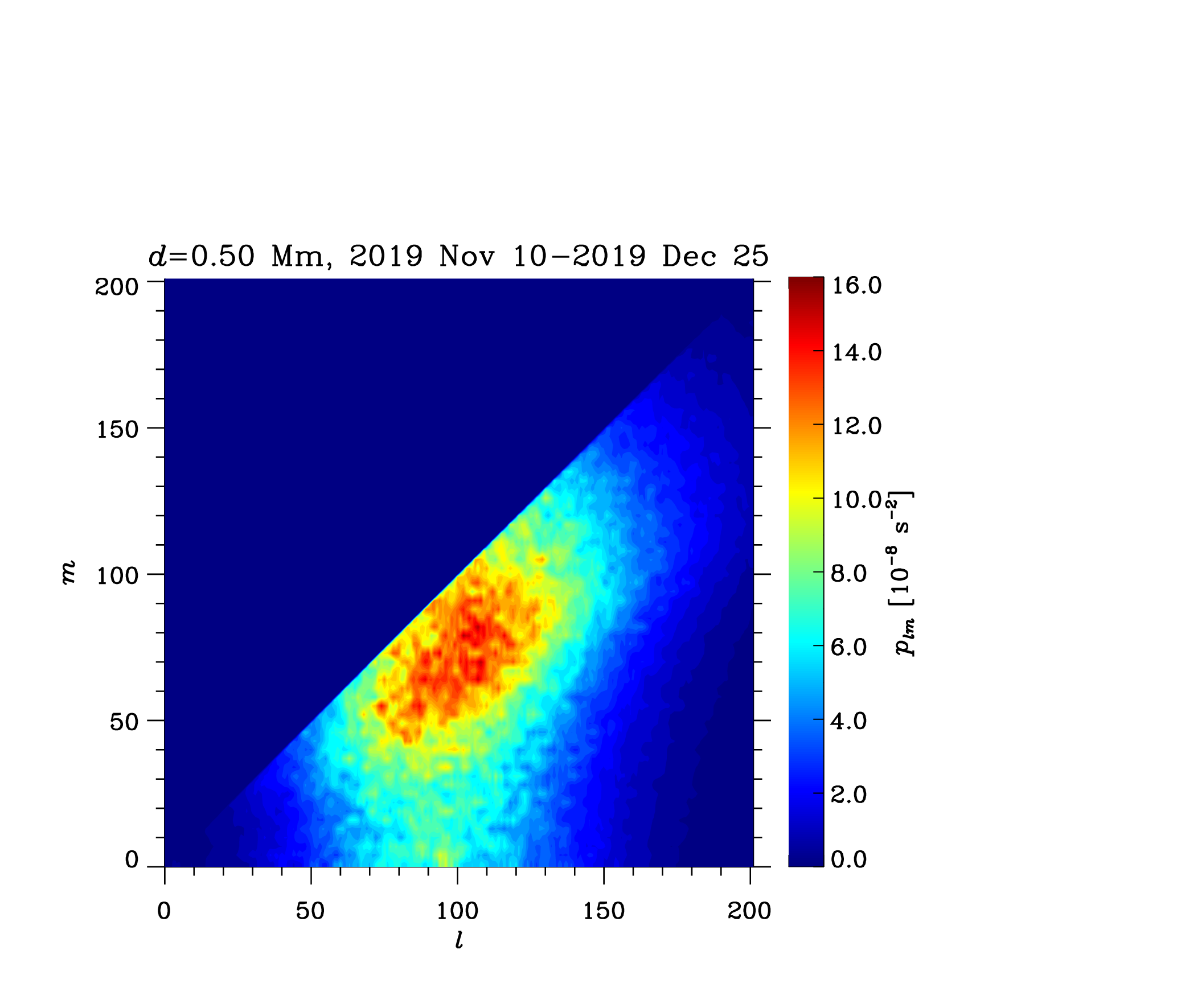}
        \includegraphics[width=0.45\textwidth,bb=20 0 780 650, clip]{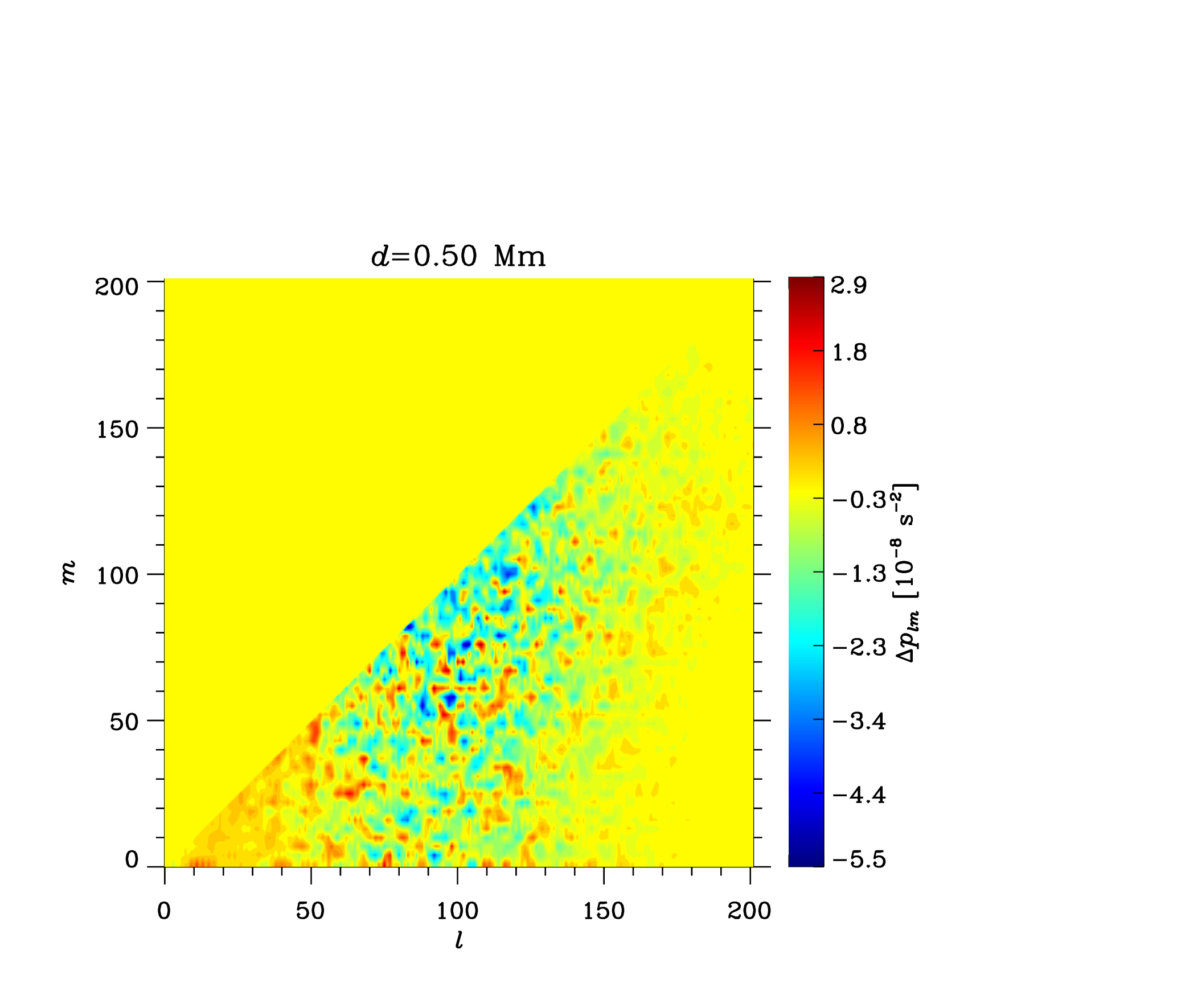}\\
        \includegraphics[width=0.45\textwidth,bb=20 0 780 650, clip]{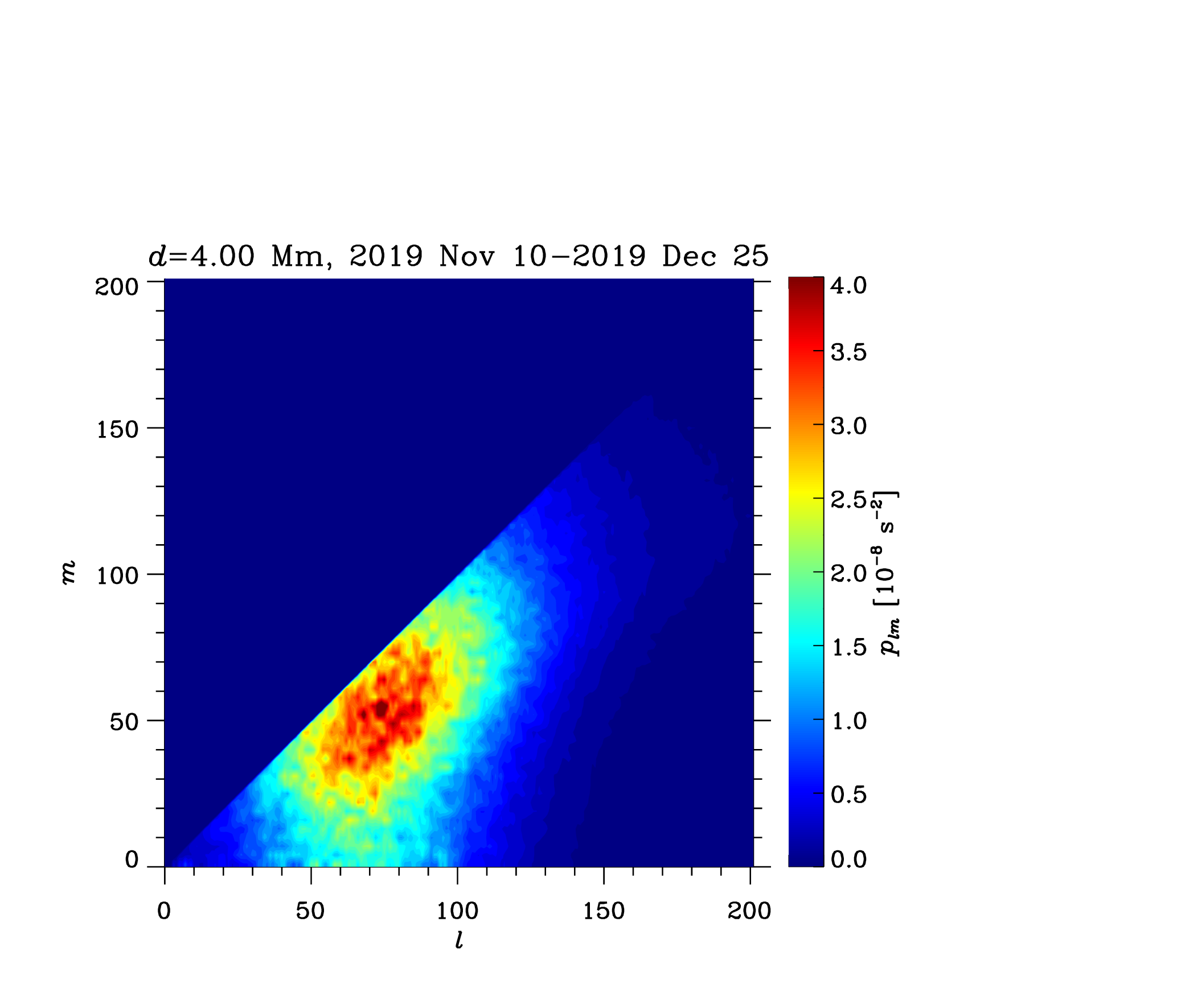}
        \includegraphics[width=0.45\textwidth,bb=20 0 780 650, clip]{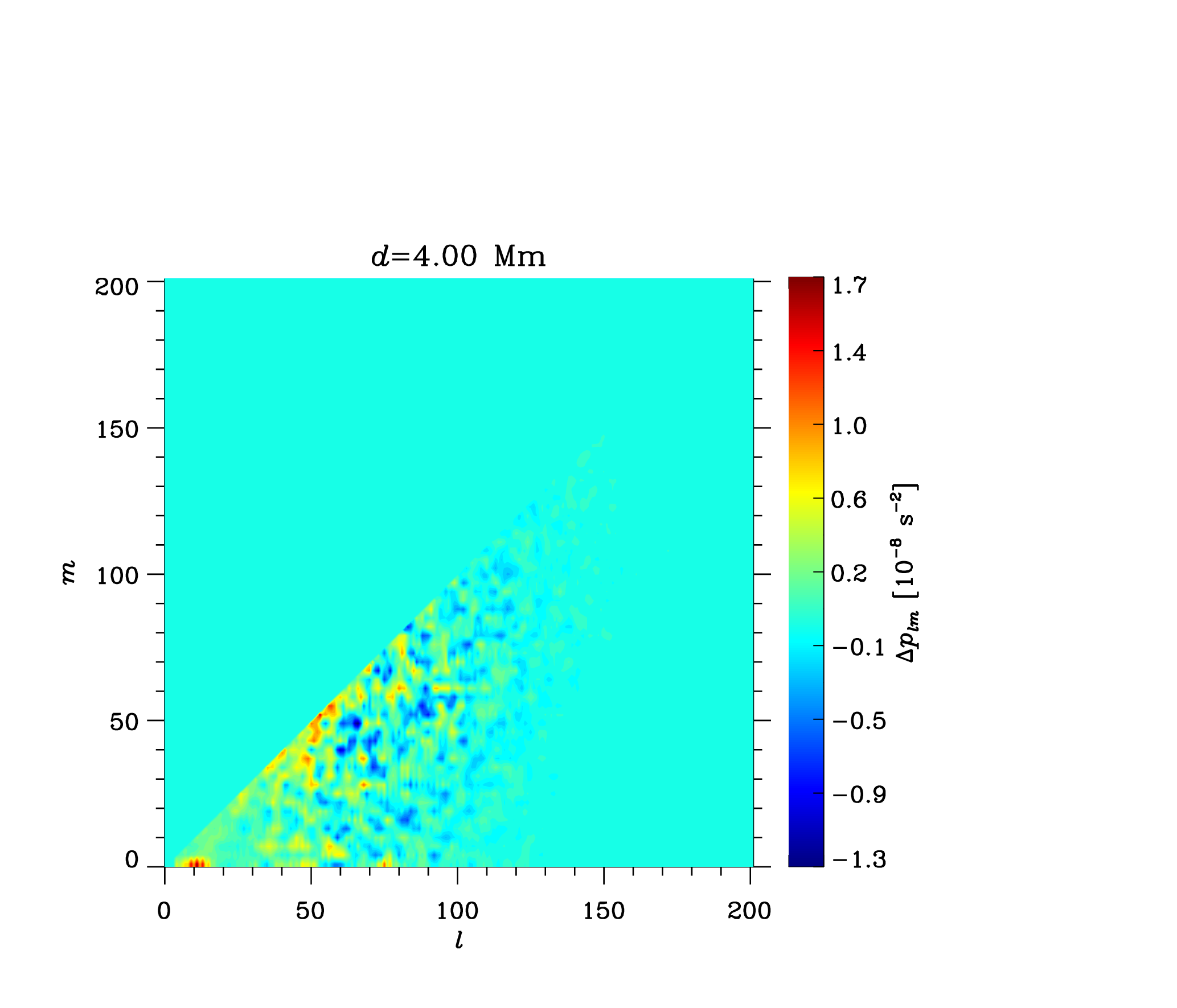}\\
        \includegraphics[width=0.45\textwidth,bb=20 0 780 650, clip]{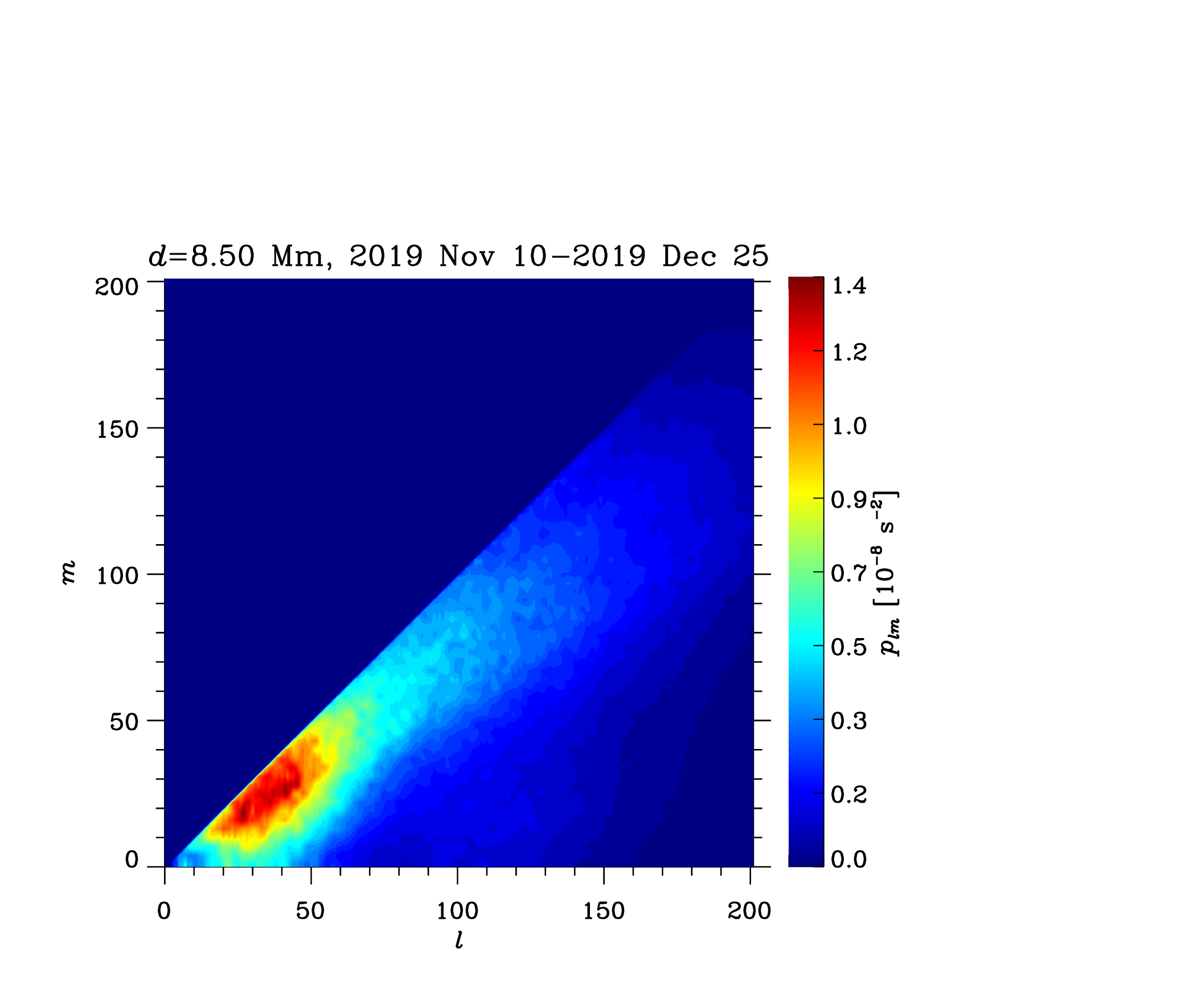}
        \includegraphics[width=0.45\textwidth,bb=20 0 780 650, clip]{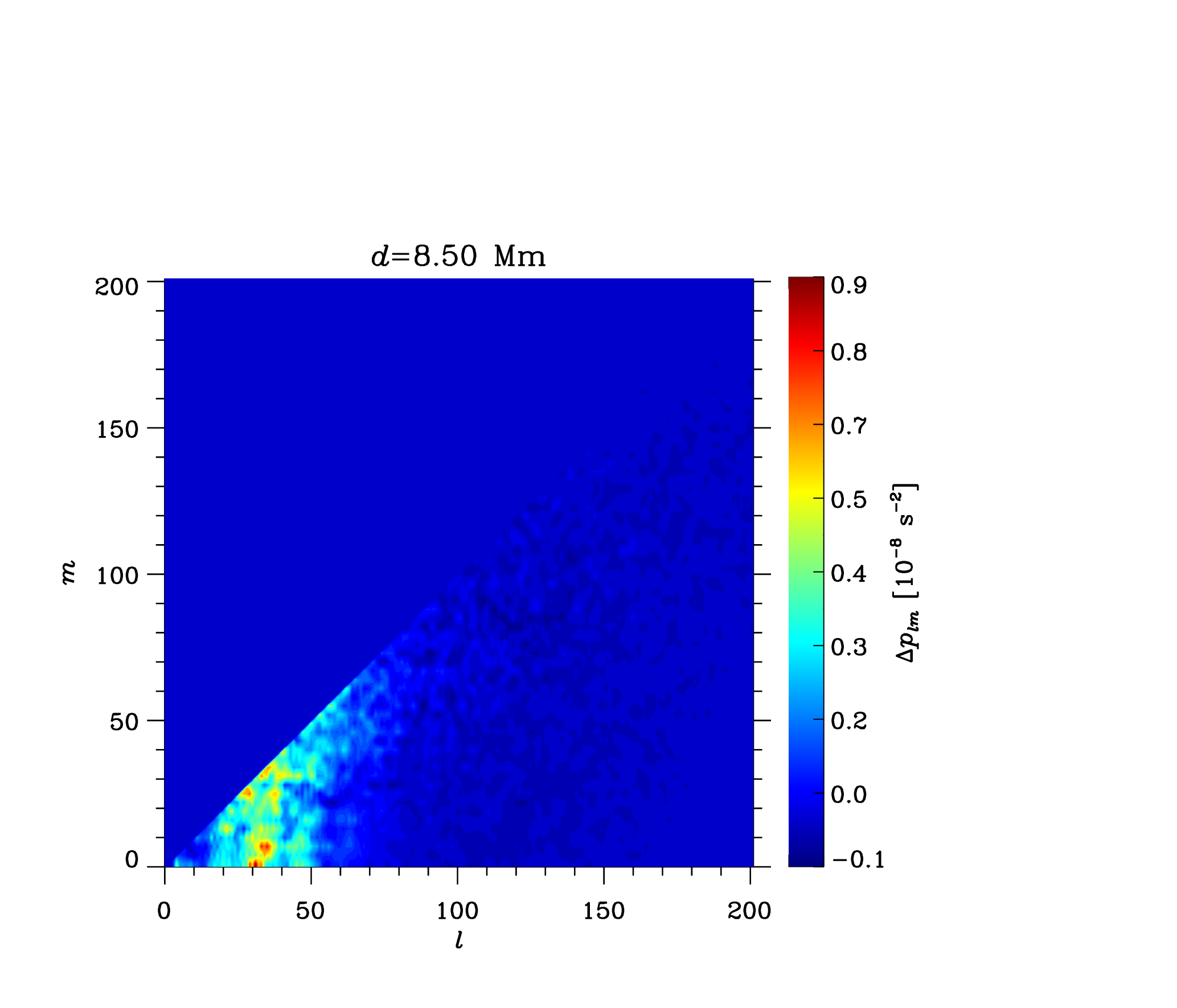}\\
        \includegraphics[width=0.45\textwidth,bb=20 0 780 650, clip]{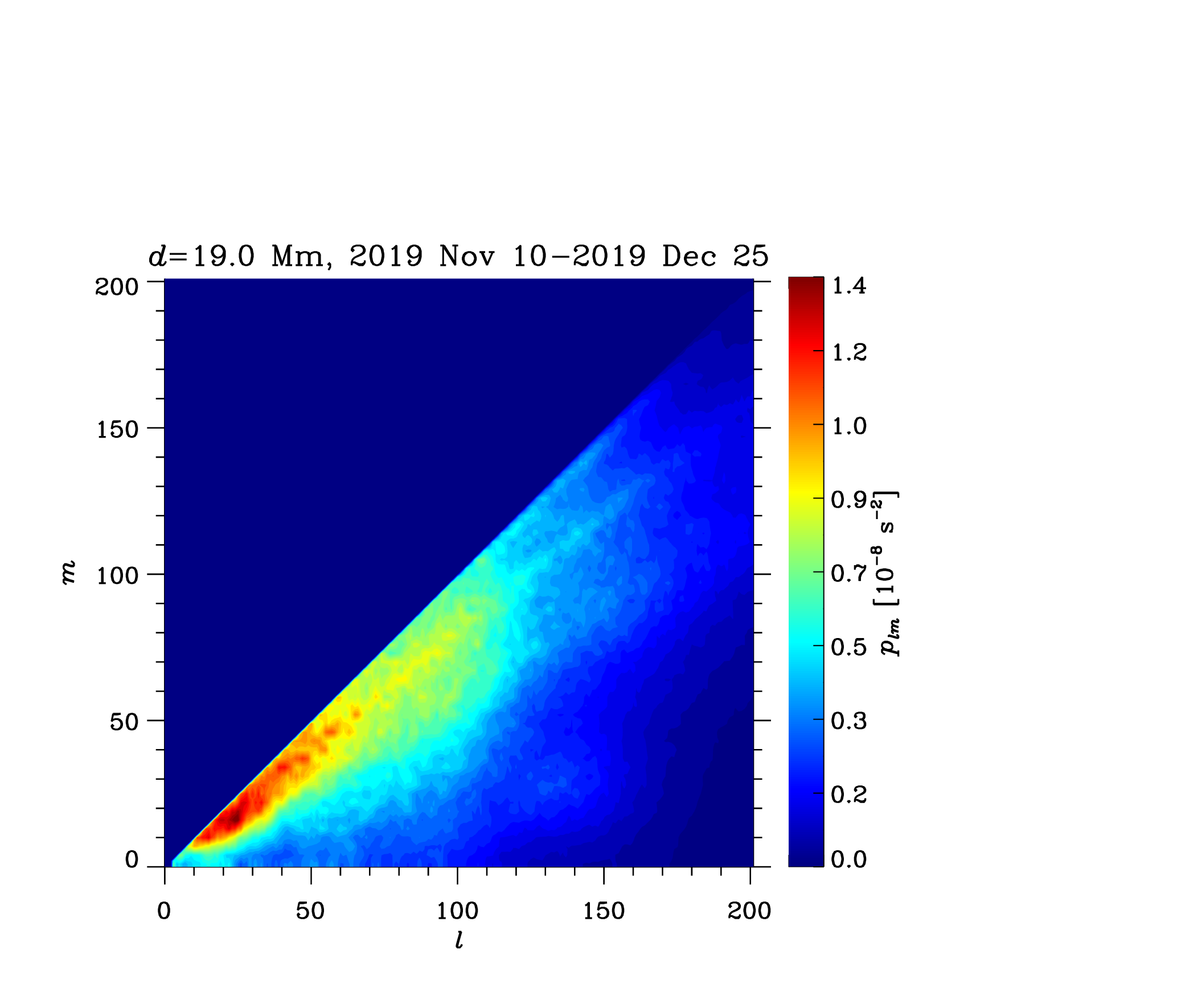}
        \includegraphics[width=0.45\textwidth,bb=20 0 780 650, clip]{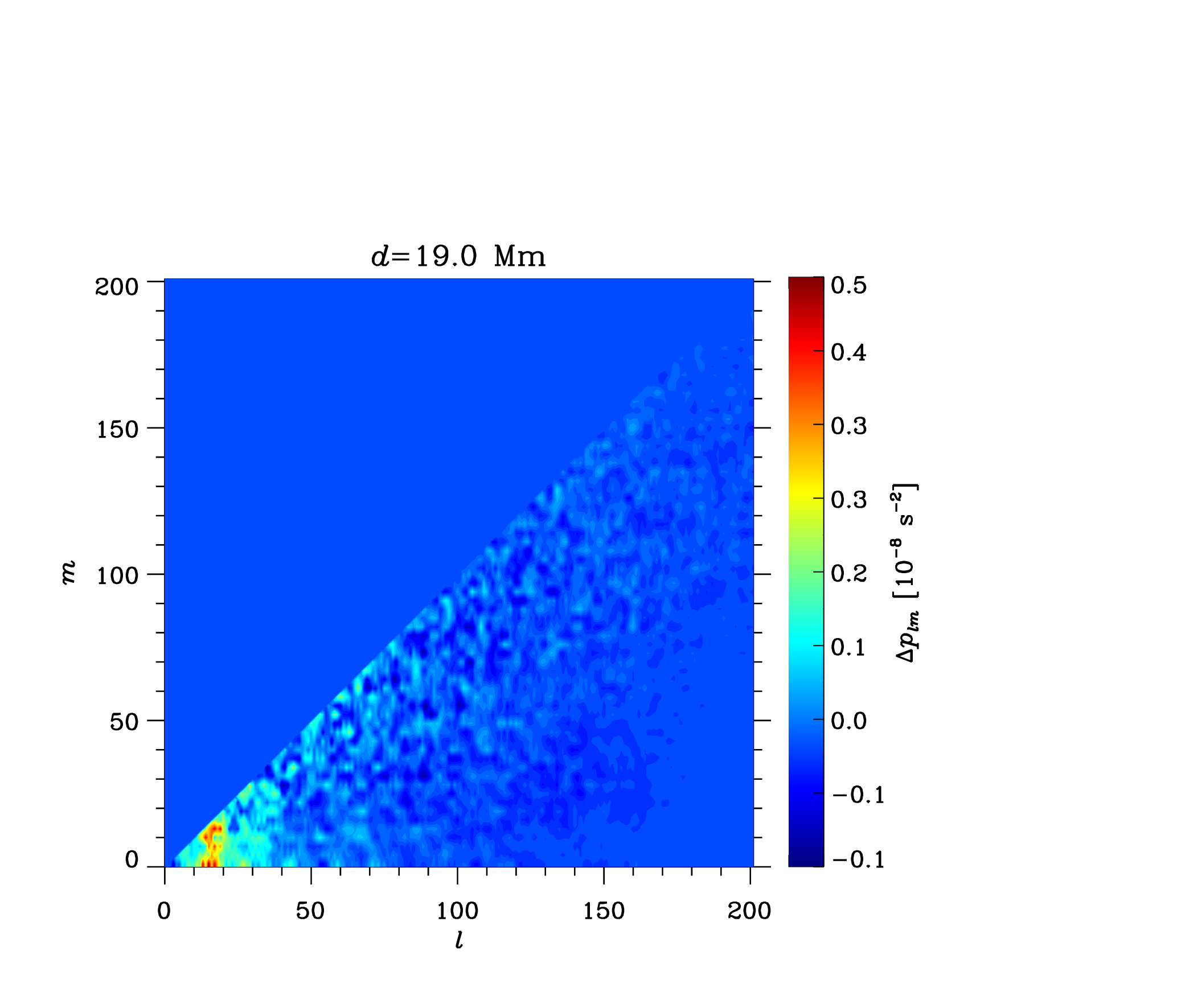}
		\caption{Left column: depth variation of a typical 45-day averaged power spectrum of the divergence field of horizontal velocity, $p_{\ell m}$, obtained for the period of low activity (9 November to 24 December 2019). The depth values and the averaging interval are indicated at the top of each panel. Right column: power-spectrum differences, $\Delta p_{\ell m}$, between the periods of high activity (9 January to 23 February 2014) and low activity (9 November to 24 December 2019) for the same depths.}\label{spectra}}
        \end{figure*}
        
Granulation is not resolved by time--distance helioseismology. To remove noise due to granulation and smaller-scale motions, we smooth the inferred velocity field with a 17.5-Mm window. We assume $\ell_{\max}=201$ as the upper limit of the spherical-harmonic expansion, corresponding to the shortest convection scales in our spectral analysis.
    
    Each spectrum occupies an area on the $(\ell,m)$ plane whose width decreases with depth, $d$, this area also shifting as a whole toward low $\ell$ values (Figure~\ref{spectra}); we have previously given a detailed discussion of these effects \citep{Getling_Kosovichev_2022}. We are now interested in quantifying the characteristic scale of convection structures at a given depth. We define the characteristic scale by the spherical-harmonic angular degree, $\ell_c$, calculated as a `center of mass' of the spectral area using the formula	
\begin{equation}
        \ell_\mathrm c =  \frac{\displaystyle\sum\limits_{\ell=1}^{\ell_\mathrm{max}} \sum\limits_{m=1}^{\ell}\ell p_{\ell m}}{\displaystyle\sum\limits_{\ell=1}^{\ell_\mathrm{max}} \sum\limits_{m=1}^{\ell} p_{\ell m}}.\label{lc}
\end{equation}
(Here, the subscript `c' stands for `center'.) More precisely, we use in this expression the $p_{\ell m}$ values exceeding some specifically chosen threshold level because taking into account small (noise) $p_{\ell m}$ values prevents the identification of the characteristic scales and blurs the regularities of their behavior. If this threshold is too low, $\ell_\mathrm c$ is not definitely correlated with the activity level, being affected by noise. Varying threshold values in excess of \mbox{$0.65\max p_{\ell m}$} does not markedly influence the $\ell_\mathrm c$ values. Thus, we assume the threshold to be at a level of \mbox{$0.65\max p_{\ell m}$}, which seems to be close to an optimum in terms of making the sought-for inferences, and replace the smaller values with zero. A transition from $\ell_\mathrm c$ to the characteristic wavelength, $\lambda_\mathrm c$, can easily be done using Eq.~\ref{lc}.

The variation of the inclination angle between the solar rotational axis and the line of sight is removed by applying low-pass filtering \citep{Getling_Kosovichev_2022}. This means that the Fourier transform of the time dependence of $\ell_\mathrm c$ is multiplied by either a Butterworth or an ideal low-pass filter, 
\begin{equation}
    H(\nu)=\frac{1}{1+(\nu/\nu_\mathrm H)^{2n}} \qquad \text{or} \qquad
    H(\nu)=\left\{
\begin{aligned}
&1\quad \mathrm{if}\ \nu \leqslant \nu_\mathrm H,\\
&0\quad \mathrm{otherwise}
\end{aligned}
\right.
    \label{Btw}
\end{equation}
(where $\nu$ is the frequency of a Fourier harmonic, $\nu_\mathrm H$ is the cutoff frequency assumed to be 0.05~$\mu$Hz and $n=4$ is set), after which an inverse Fourier transform is done.	

            \begin{figure}
	\centering
    \includegraphics[width=0.45\textwidth]{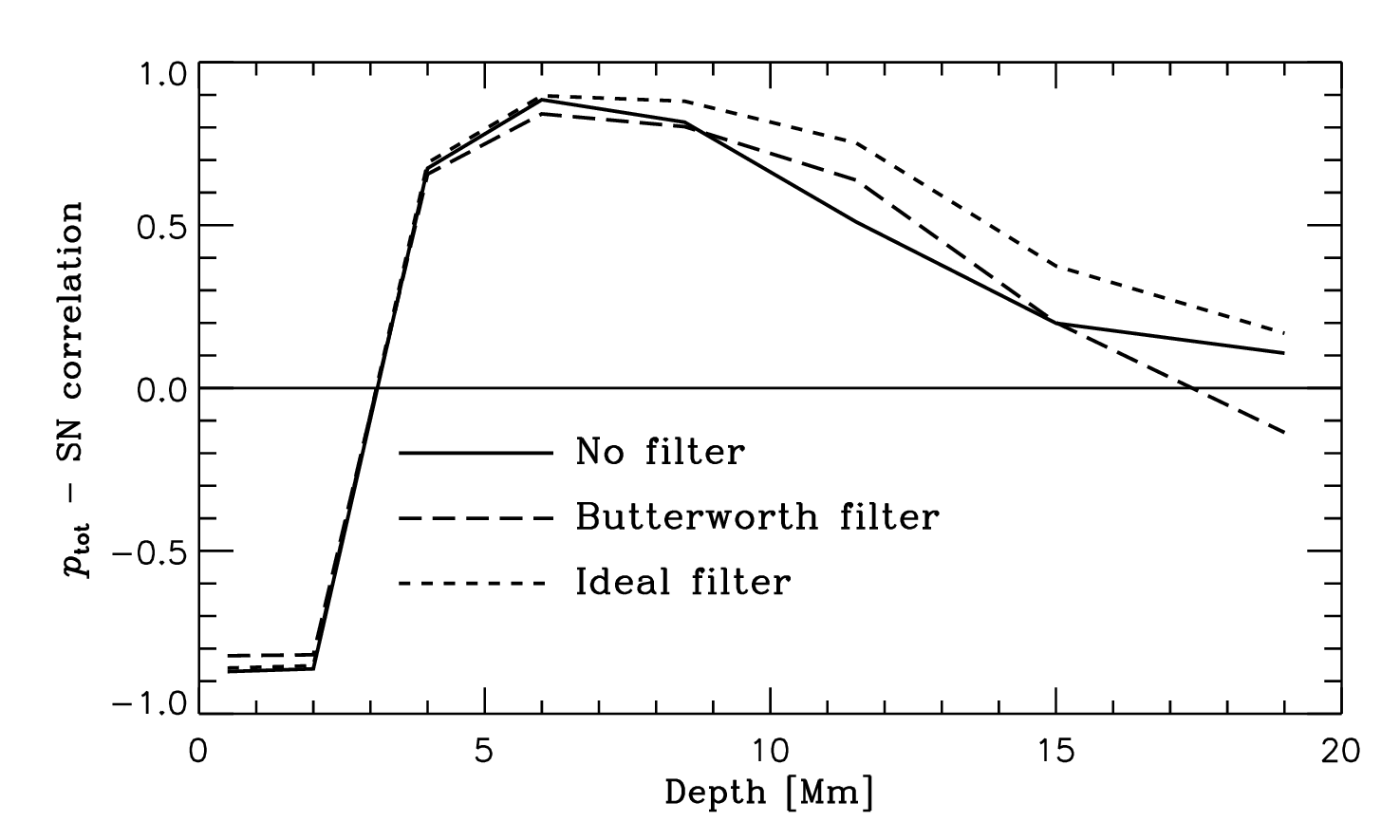}	
		\caption{Correlation of the 
        integrated spectral power of convection, $p_\mathrm{tot}$, with the sunspot number at different depths.}
			\label{int-P_corr}
	\end{figure}

   	\begin{figure*}
		\centering{
	\includegraphics[width=0.48\textwidth]{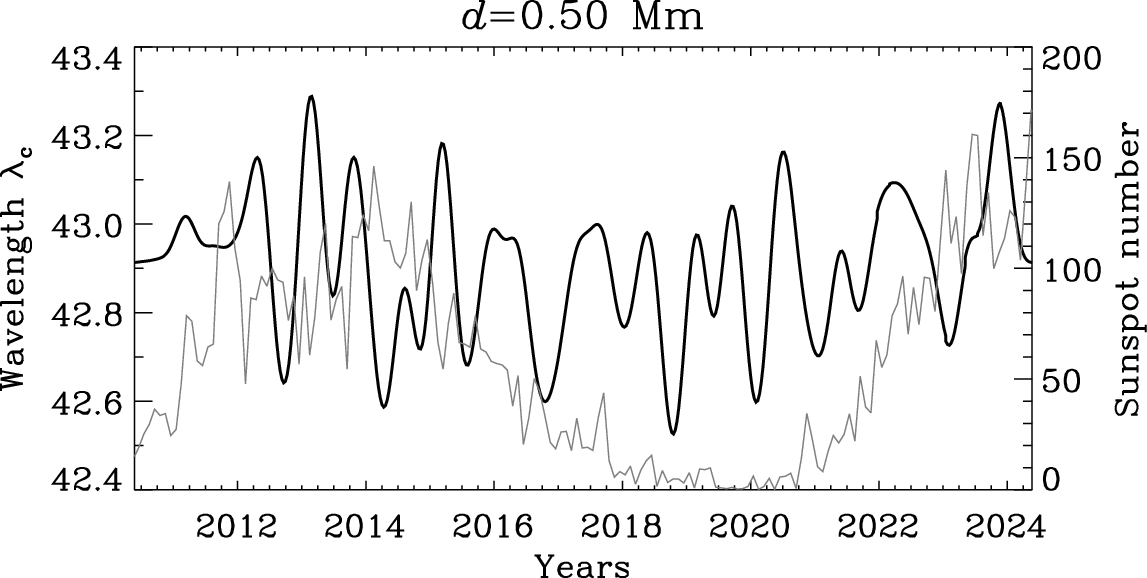}\quad	\includegraphics[width=0.48\textwidth]{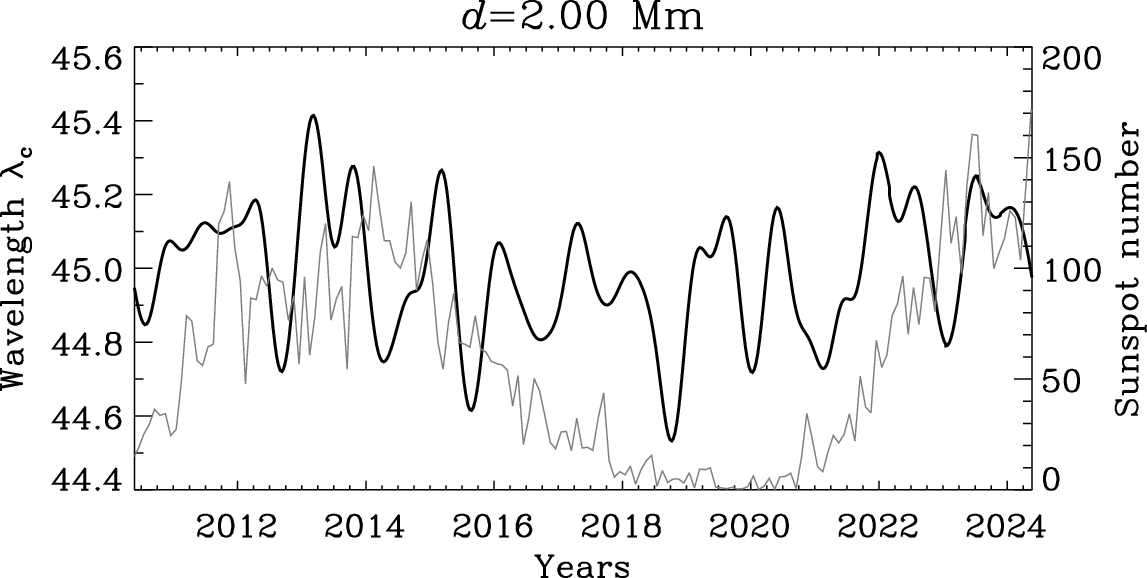}\\[12pt]
        \includegraphics[width=0.48\textwidth]{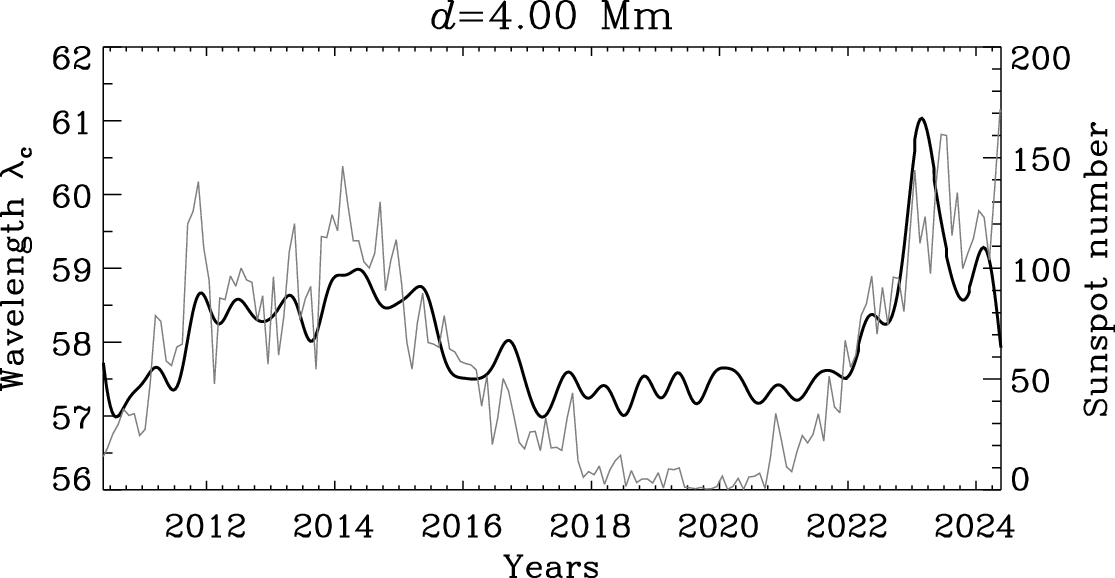}\quad\includegraphics[width=0.48\textwidth] {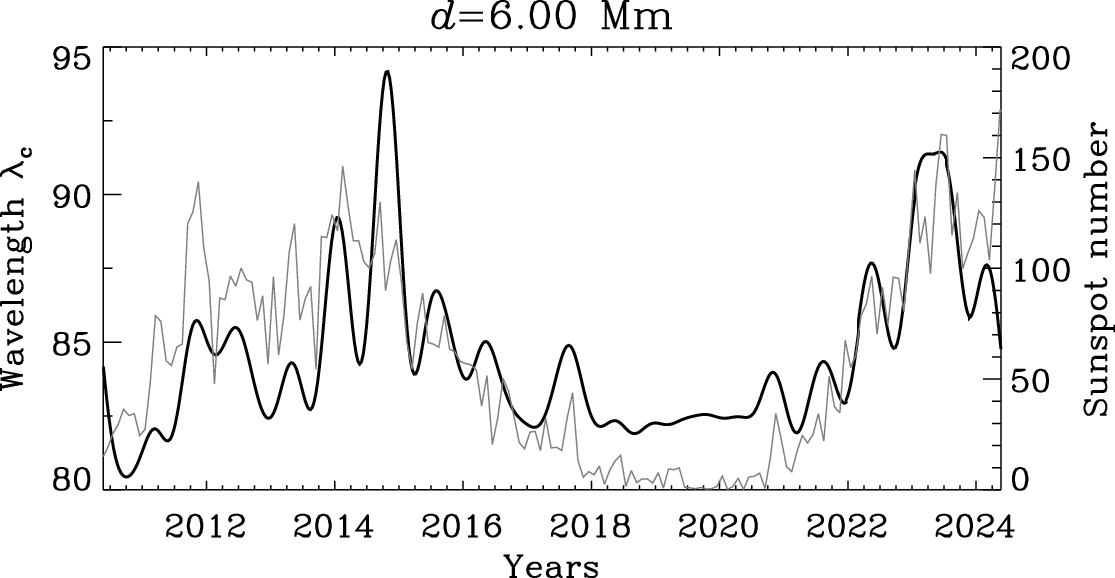}\\[12pt]
        \includegraphics[width=0.48\textwidth]{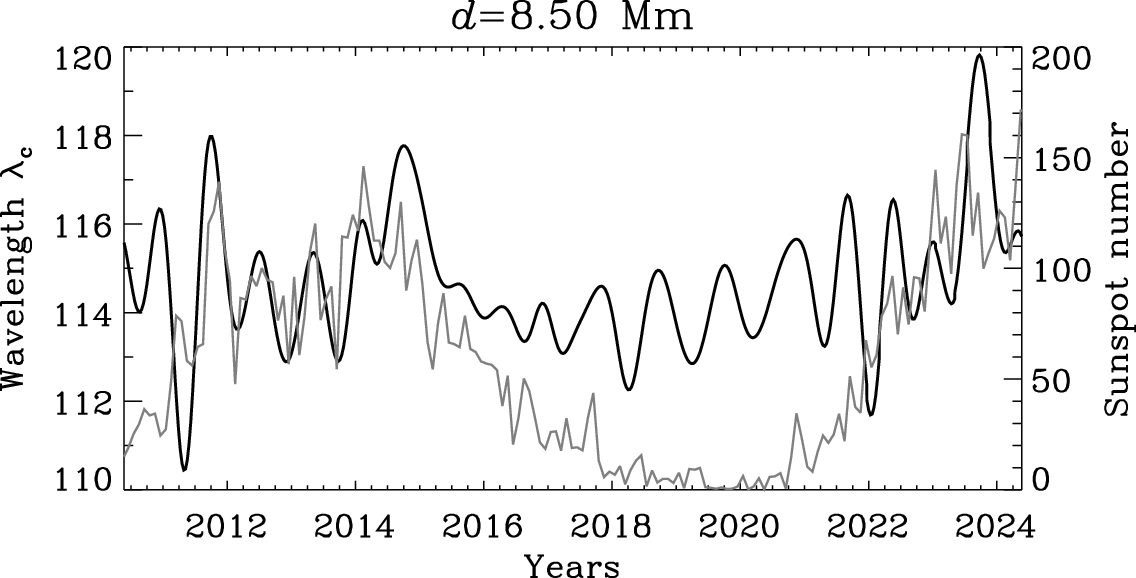}\quad\includegraphics[width=0.48\textwidth]{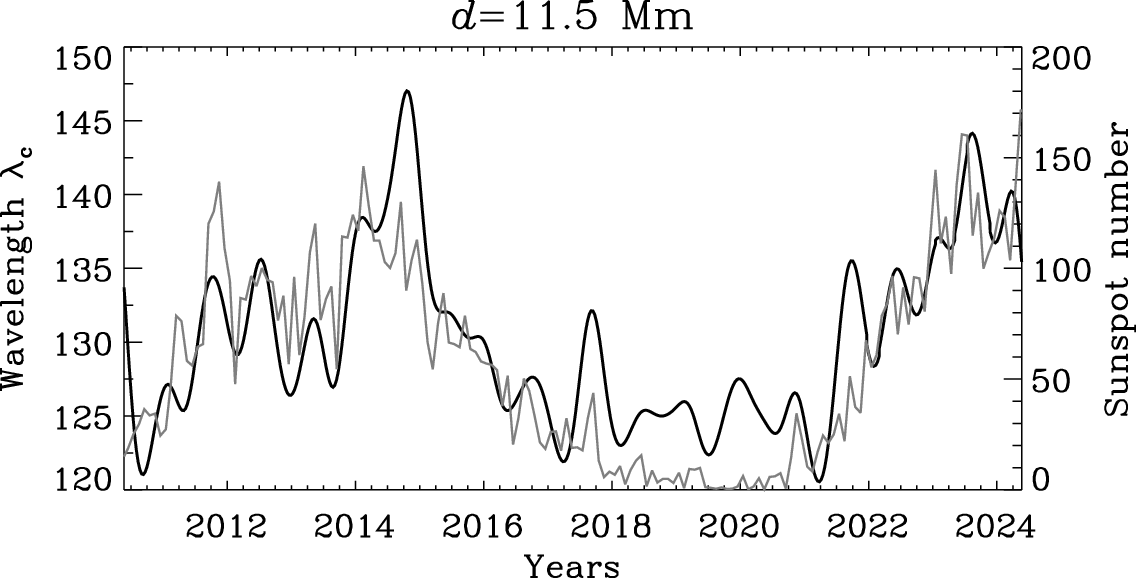}\\[12pt]
        \includegraphics[width=0.48\textwidth]{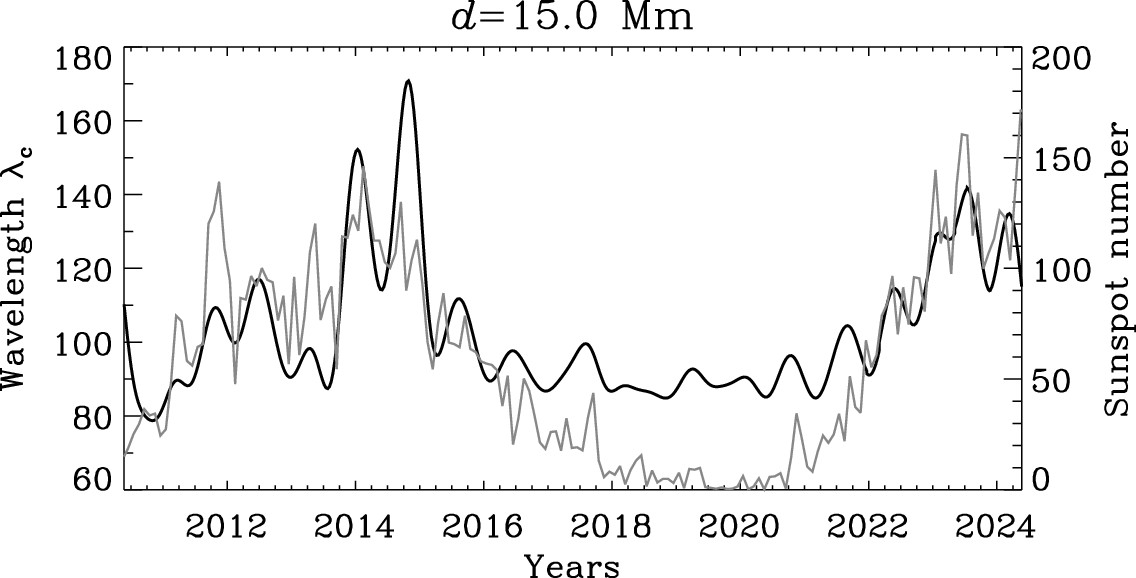}\quad\includegraphics[width=0.48\textwidth]{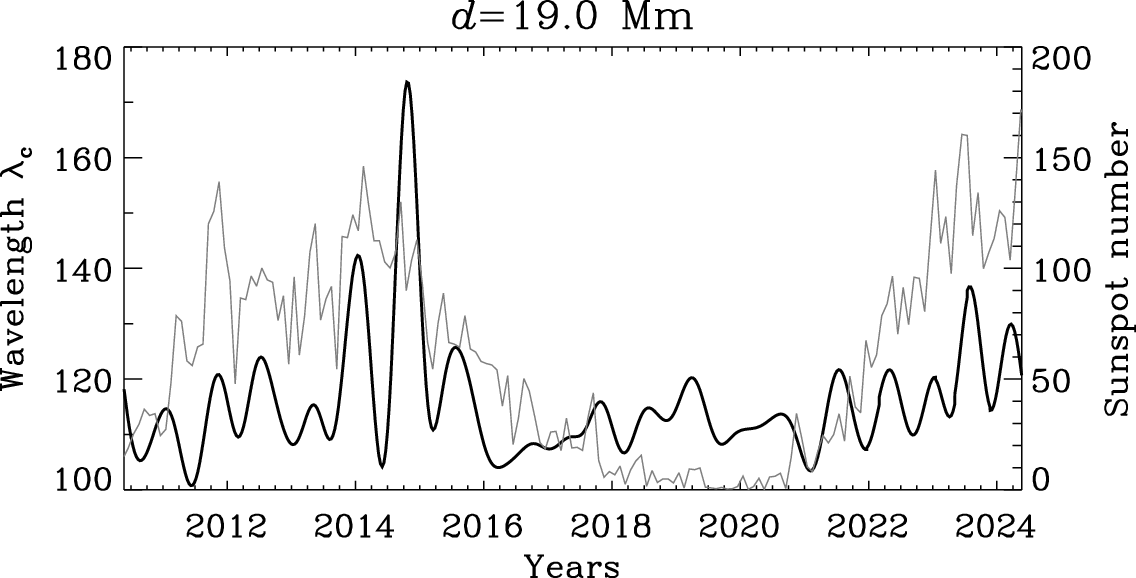}
		\caption{Time variation of the wavelength, $\lambda_\mathrm{c}$, corresponding to the Butterworth-filtered variation of the weighted mean degree of the horizontal-velocity-divergence spectrum (heavy curves). The light curves represent the variation of the sunspot number. The depth values are indicated at the top of each panel.}\label{lambdac}}
        \end{figure*}

	\begin{figure}
	\centering
    \includegraphics[width=0.45\textwidth]{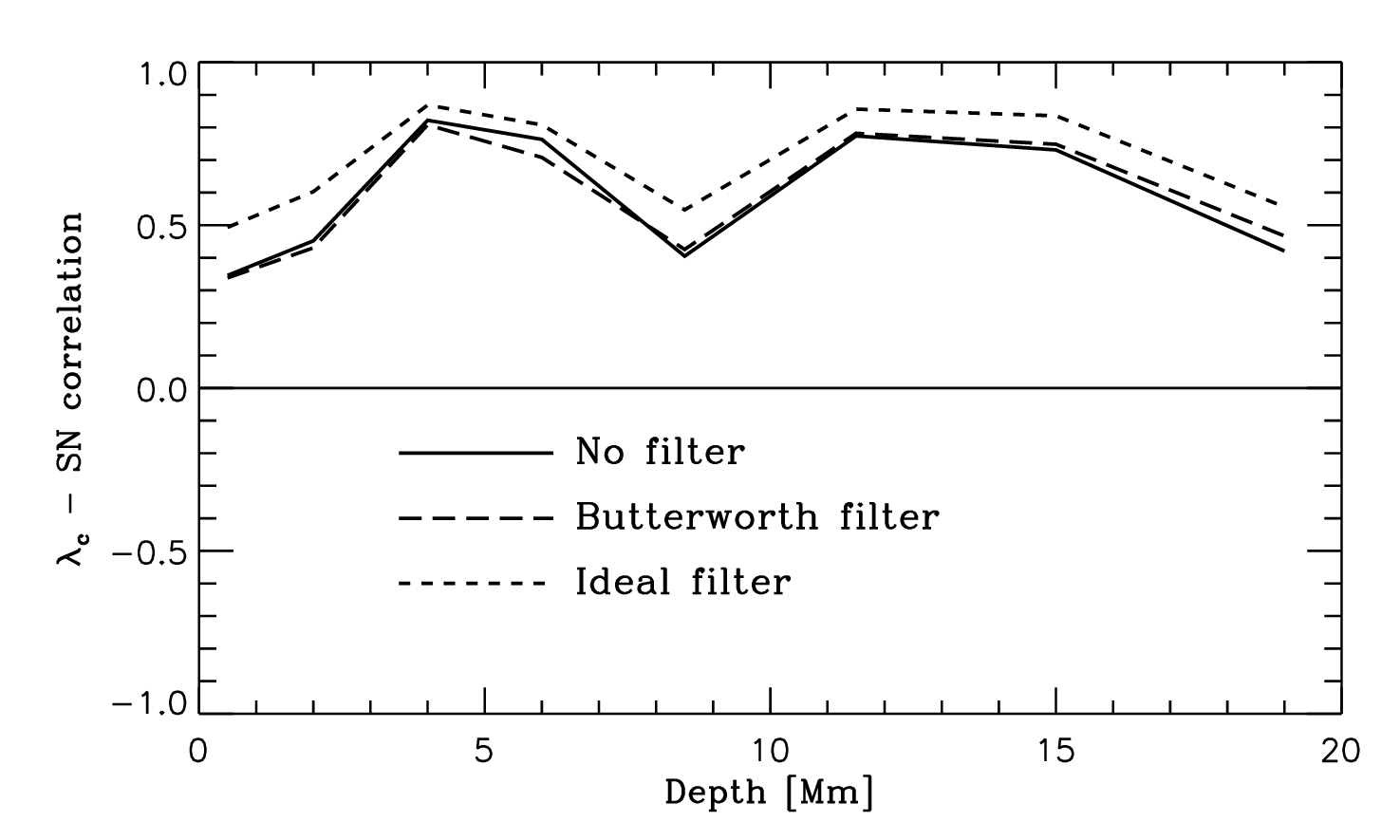}	
		\caption{Correlation of the 
        characteristic wavelength $\lambda_\mathrm c$ with the sunspot number at different depths.}
			\label{lambdacorr}
	\end{figure}

            \begin{figure}
		\centering	    
        \includegraphics[width=0.32\textwidth,bb=20 40 480 990,clip]{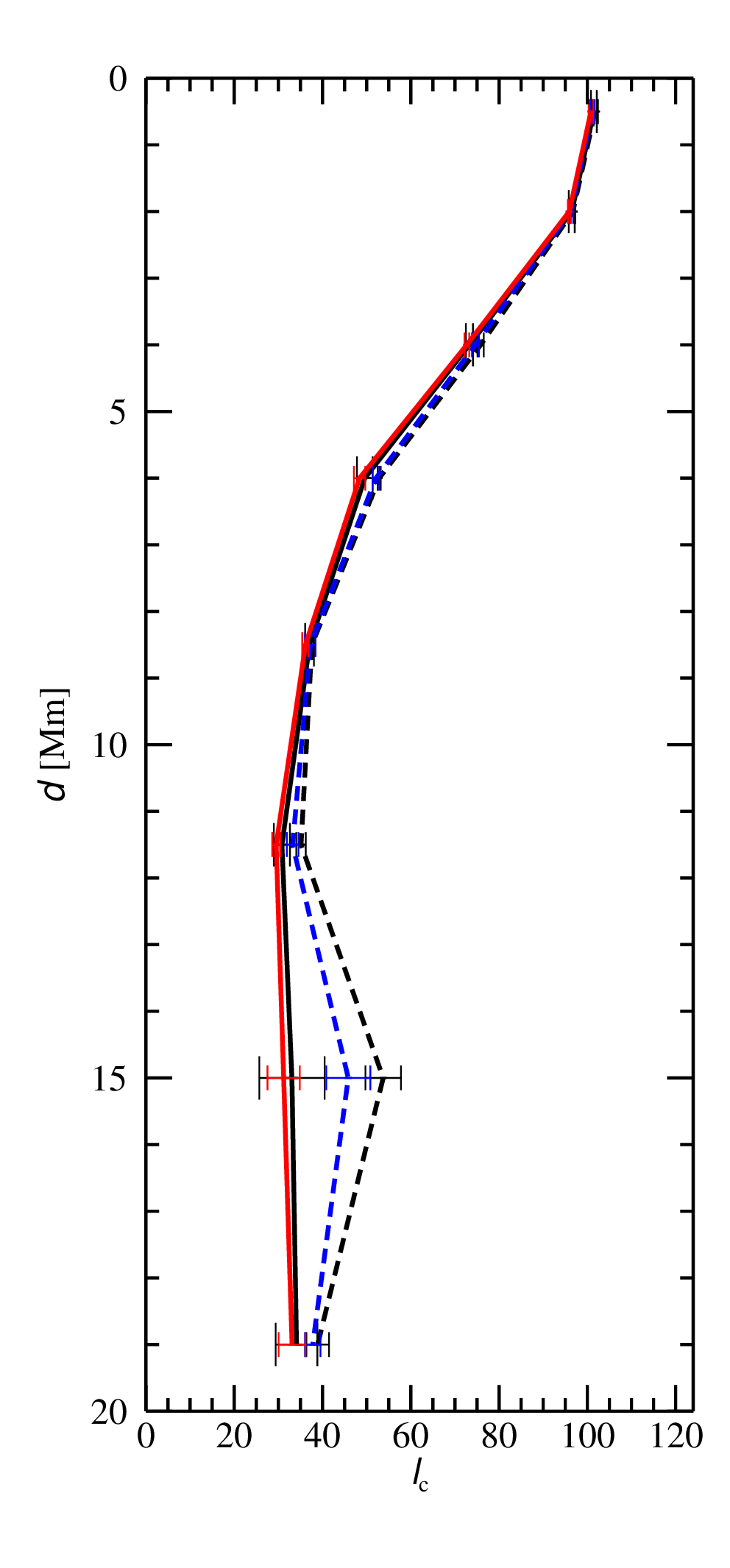}        
        \includegraphics[width=0.32\textwidth,bb=20 40 480 990,clip]{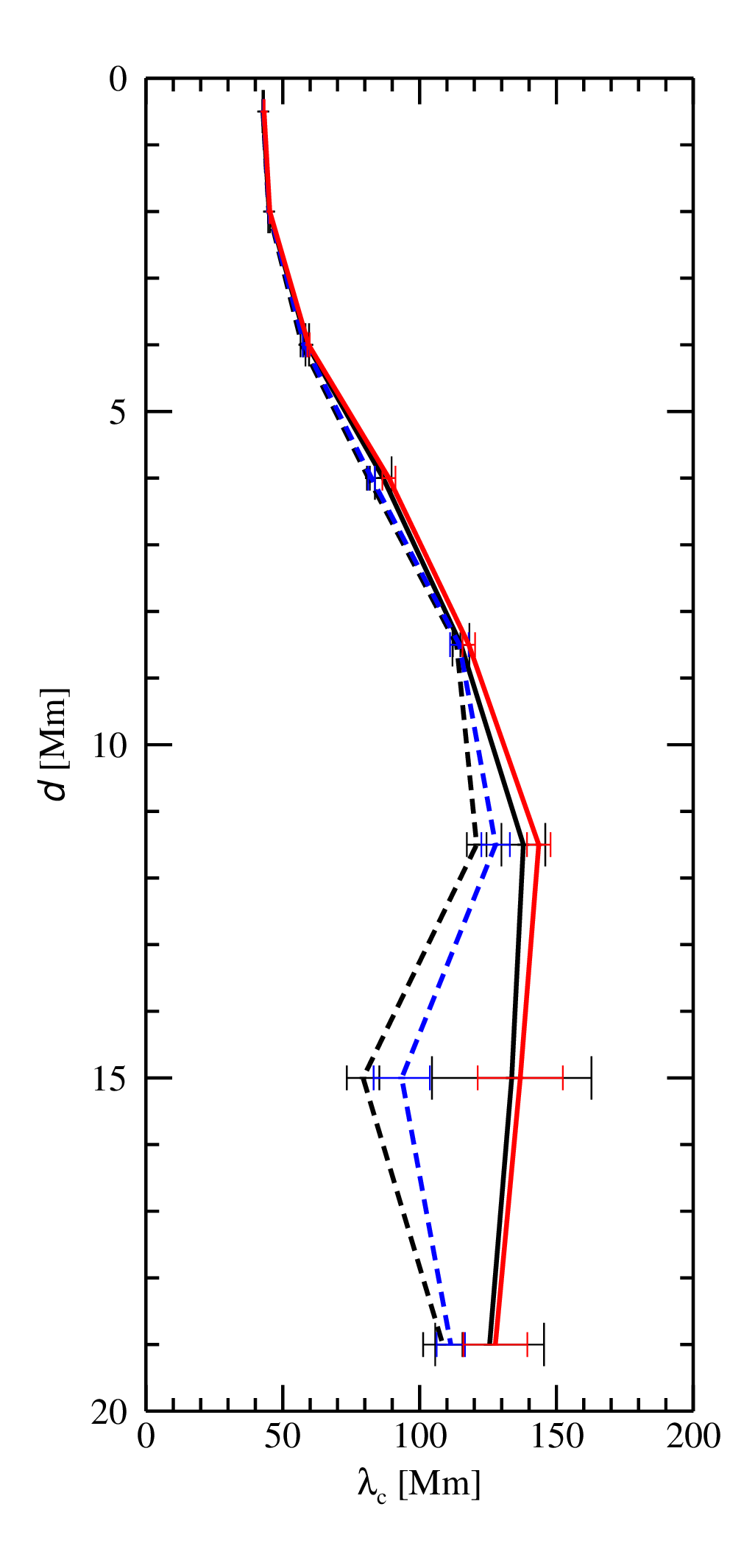}
		\caption{Depth variation of the weighted mean spherical-harmonic degree, $l_\mathrm c$ (left), and characteristic wavelength, $\lambda_\mathrm c$, inferred from $l_\mathrm c$ (right), both averaged over 360 days. Solid curves: high-activity periods centered at February 2014 (black) and immediately preceding August 2024 (red); dashed curves: low-activity periods centered at October 2019 (black) and starting from May 2010 (blue). The error bars are plotted in the same color as the corresponding mean-value curves, with error-bar endcaps being longer for the averaging period centered on February 2014.}
		\label{depthvar}
	\end{figure}
    
    \section{Results}\label{results}
	
A typical appearance of the spatial power spectra, $p_{\ell m}$, of the horizontal-velocity-field divergence at various depths, $d$, is shown in Figure~\ref{spectra}, left column, after \cite{Getling_Kosovichev_2022,Getling_Kosovichev_2025}. It can be seen that the main spectral peak narrows with the increase of $d$, moving to smaller spherical-harmonic degrees, $\ell$. While supergranular scales ($\ell \sim 70$--30, $\lambda \sim 30$--60~Mm) dominate in the upper layers, the most energetic harmonics in the deepest layers correspond to $\ell \sim 20$, with $\lambda$ broadly ranging around 300~Mm and representing giant-cell scales. A first inference concerning the time variation of the scales of convection structures can be made based on the diagrams representing the difference, $\Delta p_{\ell m}$, between the $p_{\ell m}$ distributions for the periods of high and low solar-activity levels (Figure~\ref{spectra}, right column). Specifically, structures of relatively short (supergranular) wavelengths $\lambda$ (large $\ell$ values) are weaker in the high-activity periods, especially at depths $d \sim 4$~Mm, while large-scale (giant) structures with small $\ell$ values are everywhere enhanced. It was also noted in the above-mentioned papers that the integrated power of the horizontal-velocity-divergence field is anticorrelated
with the solar-activity level in the near-surface layers and
positively correlated with it in the deeper layers (Figure~\ref{int-P_corr}). Our aim is to investigate the variations of the characteristic horizontal scale of the flow, $\lambda_\mathrm c$, with time (or activity level) and depth. 

 We use here only the Butterworth-filtered (see Eq.~\ref{Btw} in Sec.~\ref{obs}) time variation of the weighted mean degree of the horizontal-velocity-divergence spectrum, $\ell_\mathrm c$, which reveals the effects of the solar-activity level much better than the unfiltered variation does. The corresponding time variation of the characteristic wavelength, $\lambda_\mathrm c$, at different depths is shown in Figure~\ref{lambdac}. These graphs indicate that larger spatial scales of convection structures correspond to periods of higher sunspot activity. However, the degree of correlation between the sunspot activity and the convection scale changes significantly with depth. To quantify this relationship, we plot the coefficient of correlation of the wavelength, $\lambda_\mathrm c$, with the sunspot number, SN, as a function of depth in Figure~\ref{lambdacorr}. The correlation reaches its maximum values of about 0.8 at $d \approx 4$ and 11--12~Mm with a local minimum at $d \approx 8$. In our case of widely varying SN, the layers where the correlation coefficients are at minimum correspond to the least variable convection-scale values. In contrast, the highest coefficients indicate the highest variability of this scale, which is controlled by the activity level. 

 The differences in the mean spectral degrees and convection scales between different layers and their variability are illustrated by the depth dependence of the characteristic spherical-harmonic degree $\ell_\mathrm c$ and the characteristic wavelength $\lambda_\mathrm c$ averaged over 360 days (Figure~\ref{depthvar}). It can be seen that $\ell_\mathrm c$ and $\lambda_\mathrm c$ vary very little with SN in the upper layer $d \lesssim 4$~Mm. The variability slowly increases with depth to $d \approx 6$~Mm, after which it decreases to $d \approx 9$~Mm.    
 In the depth range $0 < d \lesssim 5$~Mm, the value of $\lambda_\mathrm c$ does not exceed $\sim 60$~Mm, thus lying in the range of the supergranulation scales. In these layers, it does not vary markedly during the activity cycle, being therefore weakly correlated with the sunspot number. Variations in $\lambda_\mathrm c$ are most noticeable in the depth interval of $d\gtrsim 11$, where the $\lambda_\mathrm c (d)$ profile experiences qualitative changes in the course of the solar cycle: the depth variation of $\lambda_\mathrm c$ is monotonic and weak during the periods of high activity, but has a local minimum at $d\approx 15$~Mm in the years of low activity. In this interval, the correlation coefficient reaches its second maximum and then weakly decreases over a relatively long depth interval. Everywhere in the lower half of the layer, the value of $\lambda_\mathrm c$ is in the range between $\sim 100$ and $\sim 150$~Mm, which corresponds to giant-cell scales. 

\begin{figure}
		\centering
    \includegraphics[width=0.49\textwidth,bb=50 50 772 620,clip]{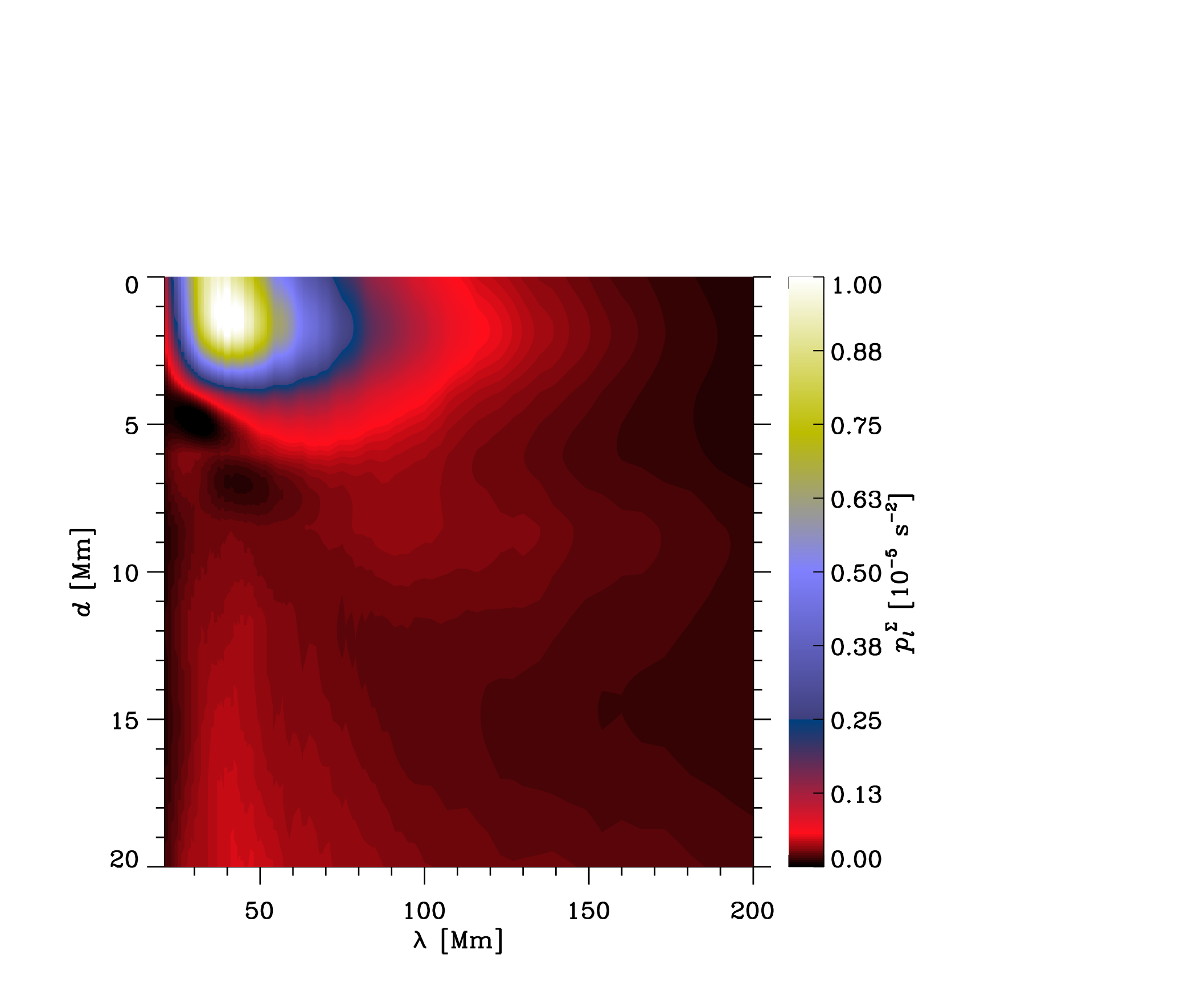}
    \includegraphics[width=0.49\textwidth,bb=50 50 772 620,clip]{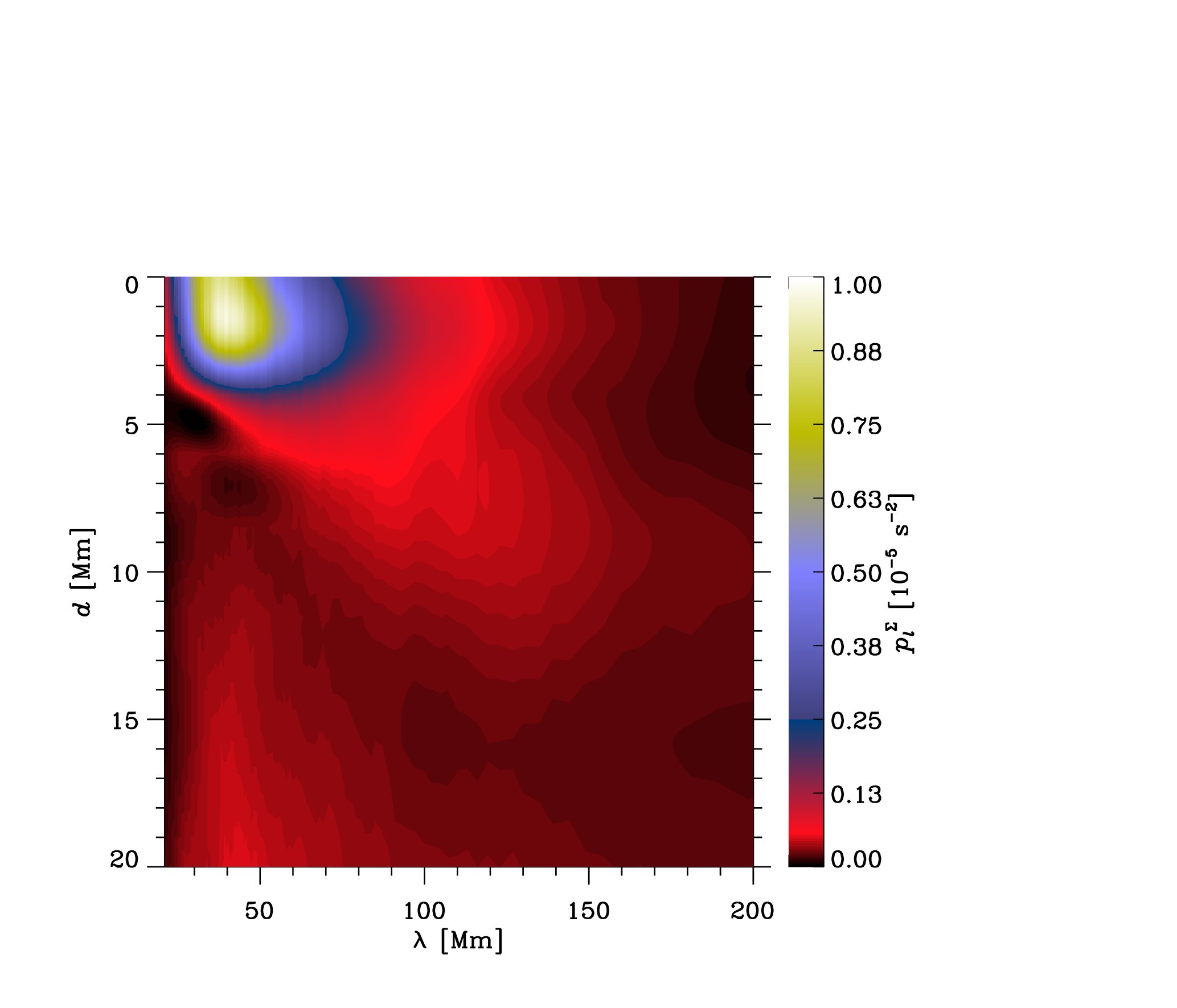}
    \caption{Distribution of the $m$-summed spectral power, $p_\ell^\Sigma$, over depth and wavelength values, obtained by 45-day averaging over the low-activity period from 9 November to 24 December 2019 (left) and the high-activity period from 9 January to 23 February 2014 (right).}
\label{m-summed_vs_depth}
    \end{figure}

     	\begin{figure}
		\centering
\includegraphics[bb=30 10 485 360,clip,width=0.4\textwidth]{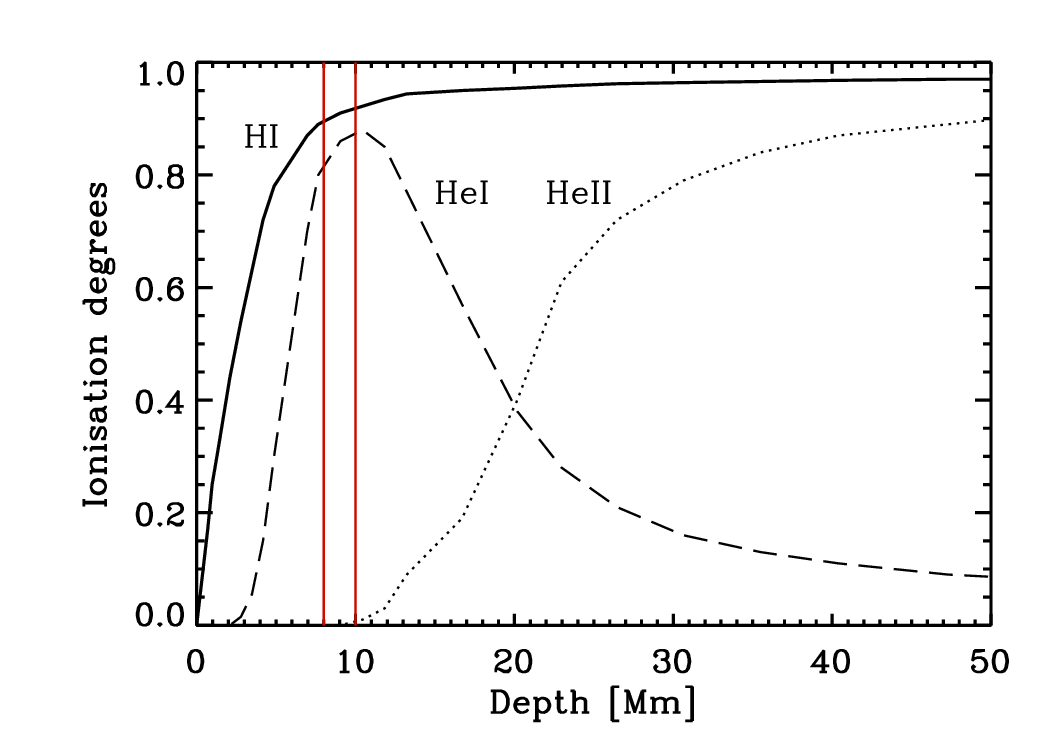}
		\caption{Ionization degrees of hydrogen and helium in the upper part of the convection zone as functions of depth. The red vertical lines mark the lower boundary of the upper and the upper boundary of the lower partial-ionization zone. The layer between them is more stable in terms of convective instability.}
		\label{ionis}
	\end{figure}

             \begin{figure}
	\centering
    \includegraphics[width=0.75\textwidth]{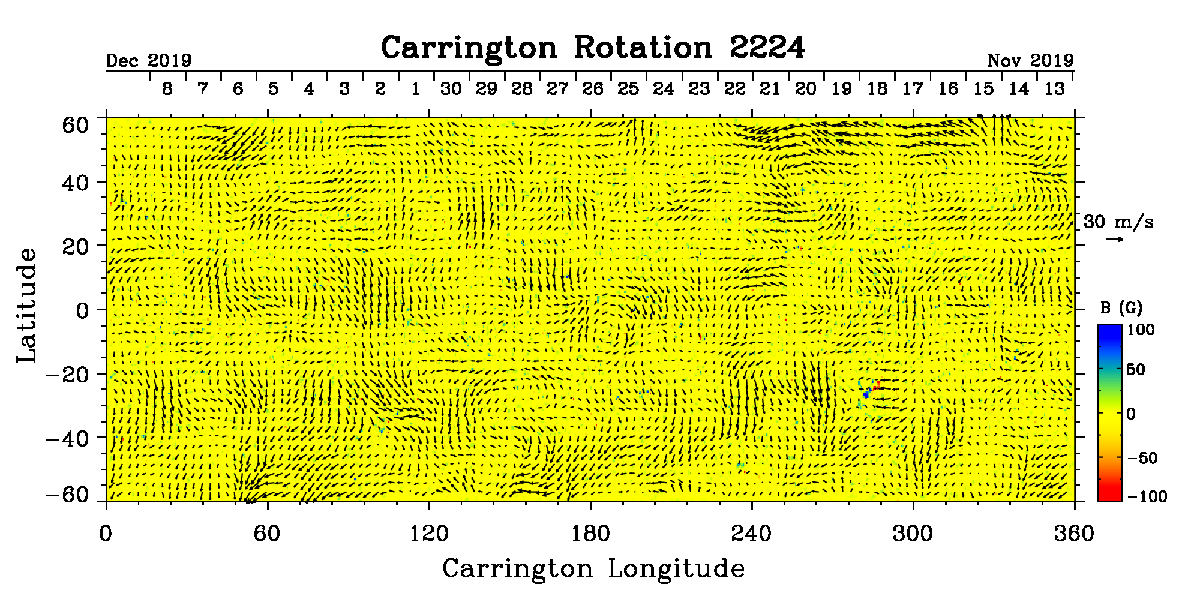}
   \includegraphics[width=0.75\textwidth]
 {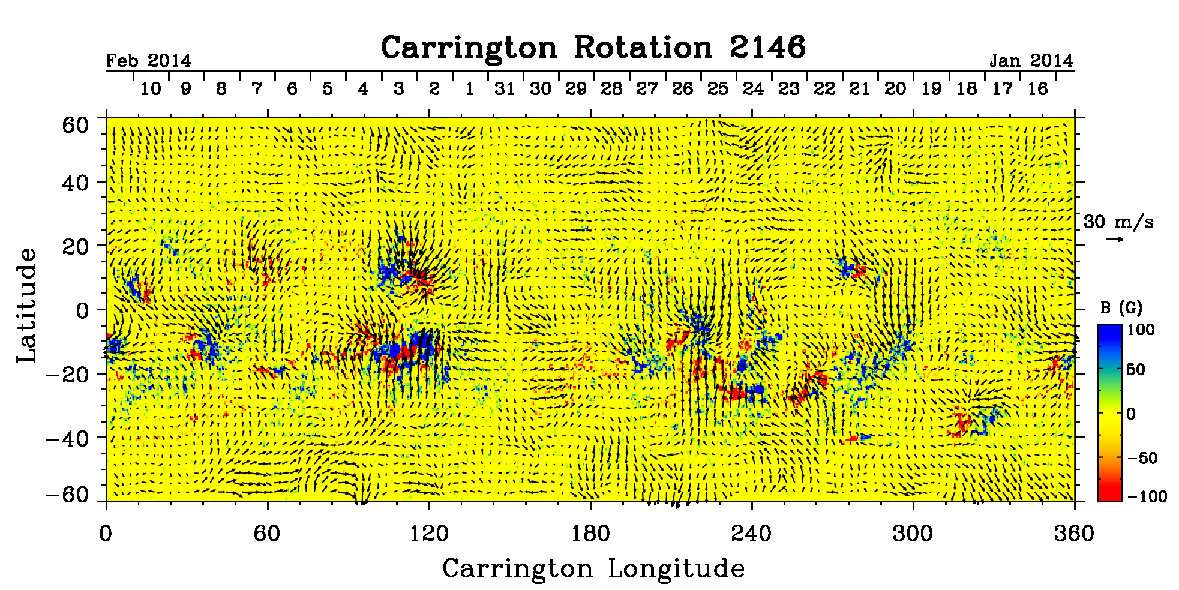}   
   \caption{Synoptic maps of horizontal subsurface flows at  depths of 0--1~Mm (arrows) and the photospheric line-of-sight magnetic field (color images) for the same periods of low (top) and high (bottom) activity from the SDO JSOC data analysis pipeline (jsoc.stanfrod.edu/data/timed/).}\label{giant_cells}
	\end{figure}
    
  Overall, the profiles of $\lambda_\mathrm c$ as a function of $d$ can be conventionally divided into three segments. The supergranular scale can easily be identified in the upper segment, $d\lesssim 4$~Mm, where supergranulation is a dominant class of structures with a stable wavelength of $\lambda_\mathrm c\sim 40$~Mm. The giant-cell scale is clearly distinguishable in the lower segment, $d\gtrsim 8$~Mm, with $\lambda_\mathrm c$ widely varying between $\sim 100$ and $\sim 150$~Mm. In the intermediate segment, which appears to represent a transition between these two, the mean scale is determined by a superposition of the two classes of structures.

Since the integrated spectral power of convection at depths $d\gtrsim 4$~Mm is positively correlated with the sunspot number, Figure~\ref{depthvar} suggests that the characteristic horizontal scale of convection structures increases with the intensification of convection. This intensification related to the enhancements of solar activity is depth-dependent, which is reflected by the two-peak appearance of the depth variation of the $\lambda_\mathrm c$--SN correlation coefficient. 

On a qualitative level, such a relationship between the intensity of convection and the characteristic scale of convection structures resembles the regularity revealed in laboratory investigations of Rayleigh--B\'enard convection by the mid-1990s: in the experiments with uncontrolled (random) initial conditions, the mean wavelength of convection structures increased with the intensification of convection, or, under experimental conditions, with the increase of the Rayleigh number [see, e.~g., comprehensive discussions of the point by \citet{ELK} and \citet{AVG}].

A local minimum of the intensity of convection at depths $d\sim 8$~Mm can be detected in diagrams representing the distribution of the $m$-summed power of harmonics of degree $\ell$ over $\lambda$ and $d$ (Figure~\ref{m-summed_vs_depth}). It is worth noting that the two peaks of correlation between $\lambda_\mathrm c$ and SN are located above and below this minimum, in depth intervals that are remarkable in terms of the ionization degrees of hydrogen and helium---in the zone of partial ionization of hydrogen and first partial ionization of helium, $d\lesssim 8$~Mm, and the zone of partial second ionization of helium, $10\lesssim d \lesssim 40$~Mm (Figure \ref{ionis}). Both these zones are strongly superadiabatic and convectively most unstable. Accordingly, the intensity of convection should be higher in these zones than at intermediate depths \citep[in particular, numerical simulations by][confirm this view]{Kitiashvili_etal_2023}. The solar-cycle variations in the convection scales may be caused by changes in the convection-zone stratification, particularly by changes of the adiabatic exponent with depth in the ionization zones, detected by helioseismology \citep{Watson2020}.  

It can be seen from Figure~\ref{m-summed_vs_depth} that the long-wavelength (100~Mm~$\lesssim \lambda_\mathrm c \lesssim $~150~Mm) part of the spectrum is especially powerful in the depth range $d\lesssim 12$~Mm during the high-activity periods. In deeper layers, the power values are considerably lower; however, as we have shown previously \citep{Getling_Kosovichev_2022,Getling_Kosovichev_2025}, the power values for these scales are comparable in the deepest and most shallow layers. We interpreted this fact to support the idea of giant cells extending to deep layers. On the other hand, the presence of the long-wavelength area of increased powers at moderate depths $d\lesssim 12$~Mm suggests that giant-scale structures may also be manifest in these layers; in this case, they could be associated with flow structures observed in the near-surface layers.

Such structures can actually be noted in the flow synoptic maps obtained from the time--distance helioseismology pipeline\footnote{jsoc.stanford.edu/data/timed} for the periods of low and high activity. The giant-scale convection structures are almost uniformly distributed within the $\pm 60\degree$ latitudinal zone covered by the data analysis (Figure~\ref{giant_cells}). They also illustrate the origin of the solar-cycle variations of such ``moderately deep''  giant-scale convection structures. During the high-activity period, the intense giant-scale flows are concentrated around active regions. Previously, large-scale flows converging around active regions were detected by the ring-diagram local helioseismology technique \citep{Haber2004,Zhao2004}. 

	\section{Conclusion}
    
We have seen that the characteristic scale of supergranulation is stable during the 11-year activity cycle. In contrast, the scale of giant cells experiences substantial variations, increasing with the enhancement of solar activity. The supergranular scale is well defined in the upper 4~Mm of the convection zone, while the giant-cell scale manifests below levels of $d \sim 8$~Mm. At depths intermediate between these two, the mean scale reflects the superposition of the two mentioned classes of structures. The power of the supergranular-scale structures reduces during the periods of high activity, while the power of the giant-scale structures increases at all depths. Near the level $d\sim 8$~Mm, where the local correlation minimum is reached, the $m$-summed power of the given-$\ell$ harmonics is also at minimum. The coefficient of correlation between the convection wavelength and the solar activity level has two peaks as a function of depth. Both of them are in the depth intervals where the spectral power of convection exhibits a substantial positive correlation with the activity level. These peaks are located in the partial-ionization zones of hydrogen and helium, which are convectively least stable; accordingly, convection is most intense in these zones. The long-wavelength (100~Mm~$\lesssim \lambda_\mathrm c \lesssim 150$~Mm) part of the spectrum exhibits enhanced power at the depths $d \lesssim 12$~Mm during the high-activity periods. This indicates that giant cells should be manifest in these layers. They may be related to the flow structures distinguishable in the synoptic maps of near-surface layers. Thus, the characteristic scale and intensity of solar convection increase with the increase of solar activity. This effect is especially pronounced in the zones of partial ionization. 

It can also be noted that the depth of $\sim 8$~Mm is close to the lower boundary of the leptocline, the layer where the rotation rate rapidly varies with depth \citep{Rozelot_etal_2025}. Therefore, the first local maximum of correlation between $\lambda_\mathrm c$ and the sunspot number is located within the leptocline. 

In this study, we treated (and termed) the revealed giant-scale structures as convection cells, although inertial waves can also be present. The question of the relationship between giant-scale convection and inertial waves was put aside. Distinguishing between them is a particular problem, which deserves a separate investigation.

\begin{acks}
The helioseismological data used here were derived from HMI observational data available, courtesy of the NASA/SDO and HMI science teams. We also used sunspot number data from the World Data Center for Sunspot Index, and Long-term Solar Observations (WDC-SILSO), Royal Observatory of Belgium, Brussels. A.G. conducted his study
under the state assignment of Lomonosov Moscow State University. 
\end{acks}
%

%
%
\bibliographystyle{spr-mp-sola}
\bibliography{Getling}  

\begin{thebibliography}{25}
\ifx\bisbn     \undefined \def\bisbn  #1{ISBN #1}\fi
\ifx\binits    \undefined \def\binits#1{#1}\fi
\ifx\bauthor   \undefined \def\bauthor#1{#1}\fi
\ifx\batitle   \undefined \def\batitle#1{#1}\fi
\ifx\bjtitle   \undefined \def\bjtitle#1{\textit{#1}}\fi
\ifx\bvolume   \undefined \def\bvolume#1{\textbf{#1}}\fi
\ifx\byear     \undefined \def\byear#1{#1}\fi
\ifx\bissue    \undefined \def\bissue#1{#1}\fi
\ifx\bfpage    \undefined \def\bfpage#1{#1}\fi
\ifx\blpage    \undefined \def\blpage #1{#1}\fi
\ifx\burl      \undefined \def\burl#1{\href{#1}{\textsf{URL}}}\fi
\ifx\href      \undefined \def\href#1#2{#2}\fi
\ifx\betal     \undefined \def\betal{et al.}\fi
\ifx\bctitle   \undefined \def\bctitle#1{#1}\fi
\ifx\beditor   \undefined \def\beditor#1{#1}\fi
\ifx\bbtitle   \undefined \def\bbtitle#1{\textit{#1}}\fi
\ifx\bedition  \undefined \def\bedition#1{#1}\fi
\ifx\bseriesno \undefined \def\bseriesno#1{\textbf{#1}}\fi
\ifx\blocation \undefined \def\blocation#1{#1}\fi
\ifx\bsertitle \undefined \def\bsertitle#1{\textit{#1}}\fi
\ifx\bsnm      \undefined \def\bsnm#1{#1}\fi
\ifx\bsuffix   \undefined \def\bsuffix#1{#1}\fi
\ifx\bparticle \undefined \def\bparticle#1{#1}\fi
\ifx\barticle  \undefined \def\barticle#1{}\fi
\ifx\binstitute  \undefined \def\binstitute#1{#1}\fi
\ifx\bpublisher  \undefined \def\bpublisher#1{#1}\fi
\ifx\doiurl    \undefined \def\doiurl#1{\href{#1}{DOI}}\fi
\makeatletter
\def\safeHref#1#2#3{\in@{http}{#2}\ifin@\href{#2}{#3}\else\href{#1#2}{#3}\fi}
\makeatother
\ifx\adsurl    \undefined
  \def\adsurl#1{\safeHref{https://ui.adsabs.harvard.edu/abs/}{#1}{ADS}}\fi
\ifx\arxivurl  \undefined
  \def\arxivurl#1{\safeHref{http://arxiv.org/abs/}{#1}{arXiv}}\fi
\ifx\botherref \undefined \def\botherref#1{}\fi
\ifx\url       \undefined \def\url#1{#1}\fi
\ifx\bchapter  \undefined \def\bchapter#1{}\fi
\ifx\bbook     \undefined \def\bbook#1{}\fi
\ifx\bcomment  \undefined \def\bcomment#1{#1}\fi
\ifx\oauthor   \undefined \def\oauthor#1{#1}\fi
\ifx\citeauthoryear \undefined\def \citeauthoryear#1{#1}\fi
\def\endbibitem {}
\ifx\bconflocation  \undefined \def\bconflocation#1{#1} \fi

\bibitem[\protect\citeauthoryear{{Chatterjee}, {Mandal}, and
  {Banerjee}}{2017}]{Chatterjee_etal_2017}
\begin{barticle}
\bauthor{\bsnm{{Chatterjee}}, \binits{S.}},
\bauthor{\bsnm{{Mandal}}, \binits{S.}},
\bauthor{\bsnm{{Banerjee}}, \binits{D.}}:
\byear{2017},
\batitle{Variation of Supergranule Parameters with Solar Cycles: Results from
  Century-long Kodaikanal Digitized Ca II K Data}.
\bjtitle{\apj}
\bvolume{841},
\bfpage{70}.
\doiurl{https://doi.org/10.3847/1538-4357/aa709d}.
\adsurl{2017ApJ...841...70C}.
\end{barticle}
\endbibitem

\bibitem[\protect\citeauthoryear{Getling}{1998}]{AVG}
\begin{bbook}
\bauthor{\bsnm{Getling}, \binits{A.V.}}:
\byear{1998},
\bbtitle{Rayleigh--B\'enard Convection: Structures and Dynamics},
\bpublisher{World Scientific; Russian version: Moscow: Editorial URSS, 1999},
\blocation{Singapore}.
\end{bbook}
\endbibitem

\bibitem[\protect\citeauthoryear{{Getling} and
  {Kosovichev}}{2022}]{Getling_Kosovichev_2022}
\begin{barticle}
\bauthor{\bsnm{{Getling}}, \binits{A.V.}},
\bauthor{\bsnm{{Kosovichev}}, \binits{A.G.}}:
\byear{2022},
\batitle{Spatial Scales and Time Variation of Solar Subsurface Convection}.
\bjtitle{\apj}
\bvolume{937},
\bfpage{41}.
\doiurl{https://doi.org/10.3847/1538-4357/ac8870}.
\adsurl{2022ApJ...937...41G}.
\end{barticle}
\endbibitem

\bibitem[\protect\citeauthoryear{{Getling} and
  {Kosovichev}}{2025}]{Getling_Kosovichev_2025}
\begin{barticle}
\bauthor{\bsnm{{Getling}}, \binits{A.V.}},
\bauthor{\bsnm{{Kosovichev}}, \binits{A.G.}}:
\byear{2025},
\batitle{Variation of Global and Local Flows in the Solar Convection Zone
  during Activity Cycles 24 and 25}.
\bjtitle{Astronomy Reports}
\bvolume{69},
\bfpage{144}.
\doiurl{https://doi.org/10.1134/S1063772925701562}.
\adsurl{2025ARep...69..144G}.
\end{barticle}
\endbibitem

\bibitem[\protect\citeauthoryear{{Gizon} et~al.}{2021}]{Gizon_etal_2021}
\begin{barticle}
\bauthor{\bsnm{{Gizon}}, \binits{L.}},
\bauthor{\bsnm{{Cameron}}, \binits{R.H.}},
\bauthor{\bsnm{{Bekki}}, \binits{Y.}},
\bauthor{\bsnm{{Birch}}, \binits{A.C.}},
\bauthor{\bsnm{{Bogart}}, \binits{R.S.}},
\bauthor{\bsnm{{Brun}}, \binits{A.S.}},
\bauthor{\bsnm{{Damiani}}, \binits{C.}},
\bauthor{\bsnm{{Fournier}}, \binits{D.}},
\bauthor{\bsnm{{Hyest}}, \binits{L.}},
\bauthor{\bsnm{{Jain}}, \binits{K.}},
\bauthor{\bsnm{{Lekshmi}}, \binits{B.}},
\bauthor{\bsnm{{Liang}}, \binits{Z.-C.}},
\bauthor{\bsnm{{Proxauf}}, \binits{B.}}:
\byear{2021},
\batitle{Solar inertial modes: Observations, identification, and diagnostic
  promise}.
\bjtitle{\aap}
\bvolume{652},
\bfpage{L6}.
\doiurl{https://doi.org/10.1051/0004-6361/202141462}.
\adsurl{2021A&A...652L...6G}.
\end{barticle}
\endbibitem

\bibitem[\protect\citeauthoryear{{Haber} et~al.}{2004}]{Haber2004}
\begin{barticle}
\bauthor{\bsnm{{Haber}}, \binits{D.A.}},
\bauthor{\bsnm{{Hindman}}, \binits{B.W.}},
\bauthor{\bsnm{{Toomre}}, \binits{J.}},
\bauthor{\bsnm{{Thompson}}, \binits{M.J.}}:
\byear{2004},
\batitle{Organized Subsurface Flows near Active Regions}.
\bjtitle{\solphys}
\bvolume{220},
\bfpage{371}.
\doiurl{https://doi.org/10.1023/B:SOLA.0000031405.52911.08}.
\adsurl{2004SoPh..220..371H}.
\end{barticle}
\endbibitem

\bibitem[\protect\citeauthoryear{{Hanson} and
  {Hanasoge}}{2024}]{Hanson_Hanasoge_2024}
\begin{barticle}
\bauthor{\bsnm{{Hanson}}, \binits{C.S.}},
\bauthor{\bsnm{{Hanasoge}}, \binits{S.}}:
\byear{2024},
\batitle{Existence of small-scale Rossby waves points to low convective
  velocity amplitudes in the Sun}.
\bjtitle{Physics of Fluids}
\bvolume{36},
\bfpage{086626}.
\doiurl{https://doi.org/10.1063/5.0216403}.
\adsurl{2024PhFl...36h6626H}.
\end{barticle}
\endbibitem

\bibitem[\protect\citeauthoryear{{Hathaway} and
  {Upton}}{2021}]{Hathaway_Upton_2021}
\begin{barticle}
\bauthor{\bsnm{{Hathaway}}, \binits{D.H.}},
\bauthor{\bsnm{{Upton}}, \binits{L.A.}}:
\byear{2021},
\batitle{Hydrodynamic Properties of the Sun's Giant Cellular Flows}.
\bjtitle{\apj}
\bvolume{908},
\bfpage{160}.
\doiurl{https://doi.org/10.3847/1538-4357/abcbfa}.
\adsurl{2021ApJ...908..160H}.
\end{barticle}
\endbibitem

\bibitem[\protect\citeauthoryear{{Hathaway}, {Upton}, and
  {Colegrove}}{2013}]{Hathaway_etal_2013}
\begin{barticle}
\bauthor{\bsnm{{Hathaway}}, \binits{D.H.}},
\bauthor{\bsnm{{Upton}}, \binits{L.}},
\bauthor{\bsnm{{Colegrove}}, \binits{O.}}:
\byear{2013},
\batitle{Giant Convection Cells Found on the {S}un}.
\bjtitle{Science}
\bvolume{342},
\bfpage{1217}.
\doiurl{https://doi.org/10.1126/science.1244682}.
\adsurl{2013Sci...342.1217H}.
\end{barticle}
\endbibitem

\bibitem[\protect\citeauthoryear{{Jeans}}{1923}]{Jeans_1923}
\begin{barticle}
\bauthor{\bsnm{{Jeans}}, \binits{J.H.}}:
\byear{1923},
\batitle{The Propagation of Earthquake Waves}.
\bjtitle{Proc. Roy. Soc. London Ser. A}
\bvolume{102},
\bfpage{554}.
\doiurl{https://doi.org/10.1098/rspa.1923.0015}.
\adsurl{1923RSPSA.102..554J}.
\end{barticle}
\endbibitem

\bibitem[\protect\citeauthoryear{{Kitiashvili}
  et~al.}{2023}]{Kitiashvili_etal_2023}
\begin{barticle}
\bauthor{\bsnm{{Kitiashvili}}, \binits{I.N.}},
\bauthor{\bsnm{{Kosovichev}}, \binits{A.G.}},
\bauthor{\bsnm{{Wray}}, \binits{A.A.}},
\bauthor{\bsnm{{Sadykov}}, \binits{V.M.}},
\bauthor{\bsnm{{Guerrero}}, \binits{G.}}:
\byear{2023},
\batitle{Leptocline as a shallow substructure of near-surface shear layer in 3D
  radiative hydrodynamic simulations}.
\bjtitle{\mnras}
\bvolume{518},
\bfpage{504}.
\doiurl{https://doi.org/10.1093/mnras/stac2946}.
\adsurl{2023MNRAS.518..504K}.
\end{barticle}
\endbibitem

\bibitem[\protect\citeauthoryear{Koschmieder}{1993}]{ELK}
\begin{bbook}
\bauthor{\bsnm{Koschmieder}, \binits{E.L.}}:
\byear{1993},
\bbtitle{B\'enard Cells and Taylor Vortices},
\bpublisher{Cambridge University Press},
\blocation{Cambridge}.
\end{bbook}
\endbibitem

\bibitem[\protect\citeauthoryear{{Mandal} and {Hanasoge}}{2024}]{Mandal2024}
\begin{barticle}
\bauthor{\bsnm{{Mandal}}, \binits{K.}},
\bauthor{\bsnm{{Hanasoge}}, \binits{S.M.}}:
\byear{2024},
\batitle{{Probing Depth Variations of Solar Inertial Modes through Normal Mode
  Coupling}}.
\bjtitle{\apj}
\bvolume{967},
\bfpage{46}.
\doiurl{https://doi.org/10.3847/1538-4357/ad391b}.
\adsurl{2024ApJ...967...46M}.
\end{barticle}
\endbibitem

\bibitem[\protect\citeauthoryear{{Mandal}, {Chatterjee}, and
  {Banerjee}}{2017}]{Mandal_etal_2017}
\begin{barticle}
\bauthor{\bsnm{{Mandal}}, \binits{S.}},
\bauthor{\bsnm{{Chatterjee}}, \binits{S.}},
\bauthor{\bsnm{{Banerjee}}, \binits{D.}}:
\byear{2017},
\batitle{Association of Supergranule Mean Scales with Solar Cycle Strengths and
  Total Solar Irradiance}.
\bjtitle{\apj}
\bvolume{844},
\bfpage{24}.
\doiurl{https://doi.org/10.3847/1538-4357/aa76e3}.
\adsurl{2017ApJ...844...24M}.
\end{barticle}
\endbibitem

\bibitem[\protect\citeauthoryear{{McIntosh} et~al.}{2011}]{McIntosh_etal_2011}
\begin{barticle}
\bauthor{\bsnm{{McIntosh}}, \binits{S.W.}},
\bauthor{\bsnm{{Leamon}}, \binits{R.J.}},
\bauthor{\bsnm{{Hock}}, \binits{R.A.}},
\bauthor{\bsnm{{Rast}}, \binits{M.P.}},
\bauthor{\bsnm{{Ulrich}}, \binits{R.K.}}:
\byear{2011},
\batitle{Observing Evolution in the Supergranular Network Length Scale During
  Periods of Low Solar Activity}.
\bjtitle{\apjl}
\bvolume{730},
\bfpage{L3}.
\doiurl{https://doi.org/10.1088/2041-8205/730/1/L3}.
\adsurl{2011ApJ...730L...3M}.
\end{barticle}
\endbibitem

\bibitem[\protect\citeauthoryear{{Papaloizou} and
  {Pringle}}{1978}]{Papaloizou_Pringle_1978}
\begin{barticle}
\bauthor{\bsnm{{Papaloizou}}, \binits{J.}},
\bauthor{\bsnm{{Pringle}}, \binits{J.E.}}:
\byear{1978},
\batitle{Non-radial oscillations of rotating stars and their relevance to the
  short-period oscillations of cataclysmic variables}.
\bjtitle{\mnras}
\bvolume{182},
\bfpage{423}.
\doiurl{https://doi.org/10.1093/mnras/182.3.423}.
\adsurl{1978MNRAS.182..423P}.
\end{barticle}
\endbibitem

\bibitem[\protect\citeauthoryear{{Raju} and {Singh}}{2002}]{Raju_Singh_2002}
\begin{barticle}
\bauthor{\bsnm{{Raju}}, \binits{K.P.}},
\bauthor{\bsnm{{Singh}}, \binits{J.}}:
\byear{2002},
\batitle{Dependence of Supergranular Length-Scales on Network Magnetic Fields}.
\bjtitle{\solphys}
\bvolume{207},
\bfpage{11}.
\doiurl{https://doi.org/10.1023/A:1015585010078}.
\adsurl{2002SoPh..207...11R}.
\end{barticle}
\endbibitem

\bibitem[\protect\citeauthoryear{{Rincon} and
  {Rieutord}}{2018}]{Rincon_Rieutord_2018}
\begin{barticle}
\bauthor{\bsnm{{Rincon}}, \binits{F.}},
\bauthor{\bsnm{{Rieutord}}, \binits{M.}}:
\byear{2018},
\batitle{The Sun's supergranulation}.
\bjtitle{Living Reviews in Solar Physics}
\bvolume{15},
\bfpage{6}.
\doiurl{https://doi.org/10.1007/s41116-018-0013-5}.
\adsurl{2018LRSP...15....6R}.
\end{barticle}
\endbibitem

\bibitem[\protect\citeauthoryear{{Rozelot}, {Kosovichev}, and
  {Kitiashvili}}{2025}]{Rozelot_etal_2025}
\begin{botherref}
\oauthor{\bsnm{{Rozelot}}, \binits{J.-P.}},
\oauthor{\bsnm{{Kosovichev}}, \binits{A.}},
\oauthor{\bsnm{{Kitiashvili}}, \binits{I.}}:
2025,
Improving Our Knowledge of the Solar Near-Surface Shear Layer: The Special Case
  of the Leptocline.
\textit{arXiv e-prints},
arXiv:2501.08021.
\adsurl{2025arXiv250108021R}.
\end{botherref}
\endbibitem

\bibitem[\protect\citeauthoryear{{Tlatov}}{2012}]{Tlatov_2012}
\begin{bchapter}
\bauthor{\bsnm{{Tlatov}}, \binits{A.G.}}:
\byear{2012},
\bctitle{Long-Term Variations of the Solar Supergranulation Size According to
  the Observations in CaIIK Line}.
In: \beditor{\bsnm{{Obridko}}, \binits{V.N.}},
\beditor{\bsnm{{Georgieva}}, \binits{K.}},
\beditor{\bsnm{{Nagovitsyn}}, \binits{Y.A.}} (eds.)
\bbtitle{The Sun: New Challenges},
\bsertitle{Astrophysics and Space Science Proceedings}
\bseriesno{30},
\bfpage{33}.
\doiurl{https://doi.org/10.1007/978-3-642-29417-4_3}.
\adsurl{2012ASSP...30...33T}.
\end{bchapter}
\endbibitem

\bibitem[\protect\citeauthoryear{{Watson} and {Basu}}{2020}]{Watson2020}
\begin{barticle}
\bauthor{\bsnm{{Watson}}, \binits{C.B.}},
\bauthor{\bsnm{{Basu}}, \binits{S.}}:
\byear{2020},
\batitle{Solar-cycle-related Changes in the Helium Ionization Zones of the
  Sun}.
\bjtitle{\apjl}
\bvolume{903},
\bfpage{L29}.
\doiurl{https://doi.org/10.3847/2041-8213/abc348}.
\adsurl{2020ApJ...903L..29W}.
\end{barticle}
\endbibitem

\bibitem[\protect\citeauthoryear{{Williams} and
  {Pesnell}}{2014}]{Williams_Pesnell_2014}
\begin{barticle}
\bauthor{\bsnm{{Williams}}, \binits{P.E.}},
\bauthor{\bsnm{{Pesnell}}, \binits{W.D.}}:
\byear{2014},
\batitle{Time-Series Analysis of Supergranule Characteristics at Solar
  Minimum}.
\bjtitle{\solphys}
\bvolume{289},
\bfpage{1101}.
\doiurl{https://doi.org/10.1007/s11207-013-0383-8}.
\adsurl{2014SoPh..289.1101W}.
\end{barticle}
\endbibitem

\bibitem[\protect\citeauthoryear{{Williams} et~al.}{2014}]{Williams_etal_2014}
\begin{barticle}
\bauthor{\bsnm{{Williams}}, \binits{P.E.}},
\bauthor{\bsnm{{Pesnell}}, \binits{W.D.}},
\bauthor{\bsnm{{Beck}}, \binits{J.G.}},
\bauthor{\bsnm{{Lee}}, \binits{S.}}:
\byear{2014},
\batitle{Analysis of Supergranule Sizes and Velocities Using Solar Dynamics
  Observatory (SDO)/ Helioseismic Magnetic Imager (HMI) and Solar and
  Heliospheric Observatory (SOHO)/ Michelson Doppler Imager (MDI)
  Dopplergrams}.
\bjtitle{\solphys}
\bvolume{289},
\bfpage{11}.
\doiurl{https://doi.org/10.1007/s11207-013-0330-8}.
\adsurl{2014SoPh..289...11W}.
\end{barticle}
\endbibitem

\bibitem[\protect\citeauthoryear{{Zhao} and {Kosovichev}}{2004}]{Zhao2004}
\begin{barticle}
\bauthor{\bsnm{{Zhao}}, \binits{J.}},
\bauthor{\bsnm{{Kosovichev}}, \binits{A.G.}}:
\byear{2004},
\batitle{Torsional Oscillation, Meridional Flows, and Vorticity Inferred in the
  Upper Convection Zone of the Sun by Time-Distance Helioseismology}.
\bjtitle{\apj}
\bvolume{603},
\bfpage{776}.
\doiurl{https://doi.org/10.1086/381489}.
\adsurl{2004ApJ...603..776Z}.
\end{barticle}
\endbibitem

\bibitem[\protect\citeauthoryear{{Zhao} et~al.}{2012}]{Zhao_2012}
\begin{barticle}
\bauthor{\bsnm{{Zhao}}, \binits{J.}},
\bauthor{\bsnm{{Couvidat}}, \binits{S.}},
\bauthor{\bsnm{{Bogart}}, \binits{R.S.}},
\bauthor{\bsnm{{Parchevsky}}, \binits{K.V.}},
\bauthor{\bsnm{{Birch}}, \binits{A.C.}},
\bauthor{\bsnm{{Duvall}}, \binits{T.L.}},
\bauthor{\bsnm{{Beck}}, \binits{J.G.}},
\bauthor{\bsnm{{Kosovichev}}, \binits{A.G.}},
\bauthor{\bsnm{{Scherrer}}, \binits{P.H.}}:
\byear{2012},
\batitle{Time-Distance Helioseismology Data-Analysis Pipeline for Helioseismic
  and Magnetic Imager Onboard Solar Dynamics Observatory (SDO/HMI) and Its
  Initial Results}.
\bjtitle{\solphys}
\bvolume{275},
\bfpage{375}.
\doiurl{https://doi.org/10.1007/s11207-011-9757-y}.
\adsurl{2012SoPh..275..375Z}.
\end{barticle}
\endbibitem

\end{thebibliography}
%
%
%
%

\end{document}